\documentclass[twocolumn]{aastex631}

\usepackage{xcolor}
\usepackage{graphicx}
\usepackage{epstopdf}
\usepackage[utf8]{inputenc}
\usepackage[T1]{fontenc}
\usepackage{url}
\usepackage{amsmath} 
\usepackage{enumitem}

\begin{document}

\def \spose#1{\hbox  to 0pt{#1\hss}}  
\newcommand{\Ha}{\hbox{{\rm H}$\alpha$}}
\newcommand{\Hb}{\hbox{{\rm H}$\beta$}}
\newcommand{\Ovi}{\hbox{{\rm O}\kern 0.1em{\sc vi}}}
\newcommand{\OIII}{\hbox{[{\rm O}\kern 0.1em{\sc iii}]}}
\newcommand{\OII}{\hbox{[{\rm O}\kern 0.1em{\sc ii}]}}
\newcommand{\lya}{Ly$\alpha$}
\newcommand{\NII}{\hbox{[{\rm N}\kern 0.1em{\sc ii}]}}
\newcommand{\SII}{\hbox{[{\rm S}\kern 0.1em{\sc ii}]}}
\newcommand{\angstrom}{\textup{\AA}}
\newcommand\ionn[2]{#1$\;${\scshape{#2}}}
\newcommand{\fesc}{$ f_{\rm esc}$}
\newcommand{\flya}{$ f_{\rm esc}^{Ly\alpha}$}
\newcommand{\kms}{\ensuremath{\mathrm{km\,s^{-1}}}}
\newcommand{\vRMS}{\ensuremath{V_{\mathrm{RMS}}}}



\title{MSA-3D: Uncovering Weak AGNs and Resolved Outflows in Disguise in $z\sim1$ Star-Forming Galaxies}

\shorttitle{MSA-3D: Uncovering Weak AGNs }

\shortauthors{Roy et al.}


\correspondingauthor{Namrata Roy}
\email{namratar@asu.edu}

\author[0000-0002-4430-8846]{Namrata Roy}
\affiliation{School of Earth and Space Exploration, Arizona State University, Tempe, AZ 85287}

\author[0000-0002-6586-4446]{Alaina Henry}
\affiliation{Space Telescope Science Institute, 3700 San Martin Drive, Baltimore, MD 21218, USA}

\author[0000-0001-5860-3419]{Tucker Jones}
\affiliation{Department of Physics and Astronomy, University of California, Davis, 1 Shields Ave, Davis, CA 95616, USA}

\author[0000-0001-6371-6274]{Ivana Bari\v{s}i\'{c}}
\affiliation{Department of Physics and Astronomy, University of California, Davis, 1 Shields Avenue, Davis, CA 95616, USA}

\author[0000-0003-4792-9119]{Ryan L. Sanders}
\affiliation{Department of Physics and Astronomy, University of Kentucky, 505 Rose Street, Lexington, KY 40506, USA}
\affiliation{Department of Physics and Astronomy, University of California, Davis, One Shields Ave, Davis, CA 95616, USA}

\author[0000-0001-9742-3138]{Kevin Bundy}
\affiliation{UCO/Lick Observatory, University of California, Santa Cruz, 1156 High Street, Santa Cruz, CA 95064, USA}

\author[0000-0001-6919-1237]{Matthew A. Malkan}
\affiliation{Department of Physics and Astronomy, University of California, 475 Portola Plaza, Los Angeles, CA 90095, USA}

\author[0000-0003-2804-0648]{Themiya Nanayakkara}
\affiliation{Centre for Astrophysics and Supercomputing, Swinburne University of Technology, Hawthorn, VIC 3122, Australia}


\author[0000-0002-3254-9044]{Karl Glazebrook}
\affiliation{Centre for Astrophysics and Supercomputing, Swinburne University of Technology, Hawthorn, VIC 3122, Australia}

\author[0000-0001-6670-6370]{Timothy Heckman}
\affiliation{The William H. Miller III Department of Physics and Astronomy, The Johns Hopkins University, Baltimore, MD 21218, USA}
\affiliation{School of Earth and Space Exploration, Arizona State University, Tempe, AZ 85287}

\author[0000-0001-6703-4676]{Juan M. Espejo Salcedo}
\affiliation{Max-Planck-Institut für extraterrestische Physik (MPE), Giessenbachstraße 1., 85748 Garching, Germany}

\author[0000-0002-9373-3865]{Xin Wang}
\affiliation{School of Astronomy and Space Science, University of Chinese Academy of Sciences (UCAS), Beijing 100049, China}
\affiliation{National Astronomical Observatories, Chinese Academy of Sciences, Beijing 100101, China}
\affiliation{Institute for Frontiers in Astronomy and Astrophysics, Beijing Normal University, Beijing 102206, China}


\author[0000-0002-1527-0762]{Danail Obreschkow}
\affiliation{International Centre for Radio Astronomy Research (ICRAR), M468, University of Western Australia, Perth, WA 6009, Australia}
\affiliation{Australian Research Council, ARC Centre of Excellence for All Sky Astrophysics in 3 Dimensions (ASTRO 3D), Australia}

\author[0000-0002-8460-0390]{Tommaso Treu}
\affiliation{Department of Physics and Astronomy, University of California, Los Angeles, CA 90095-1547, USA}


\begin{abstract}
We present spatially resolved rest-optical spectroscopy of 38 star-forming galaxies at \(0.5<z<1.5\) from the JWST/NIRSpec MSA-3D survey, which uses slit-stepping to build IFU-like datacubes at \(\sim0.1^{\prime\prime}\) resolution. We map emission-line morphology, excitation, and kinematics of the warm ionized gas using \([\mathrm{N\,II}]/\mathrm{H}\alpha\), \([\mathrm{S\,II}]/\mathrm{H}\alpha\), and \([\mathrm{O\,III}]/\mathrm{H}\beta\). Relative to \(z\sim0\) galaxies at fixed stellar mass, our sources show systematically lower \([\mathrm{N\,II}]/\mathrm{H}\alpha\) and \([\mathrm{S\,II}]/\mathrm{H}\alpha\) and elevated \([\mathrm{O\,III}]/\mathrm{H}\beta\), consistent with harder radiation fields and lower metallicities. Radially, \([\mathrm{O\,III}]/\mathrm{H}\beta\) profiles are typically flat or mildly positive, whereas \([\mathrm{N\,II}]/\mathrm{H}\alpha\) also remains flat or declines outward, mirroring metallicity trends. On kpc scales, we find a strong positive correlation between \([\mathrm{N\,II}]/\mathrm{H}\alpha\) and velocity dispersion \((\sigma)\), linking local excitation to turbulent or shock-driven kinematics. Six galaxies (\(\sim16\%\)  of the sample) host spatially localized regions  with elevated \([\mathrm{N\,II}]/\mathrm{H}\alpha\), high \(\mathrm{EW}(\mathrm{H}\alpha)\), and \(V_{\mathrm{RMS}}\equiv\sqrt{V^{2}+\sigma^{2}}>200~\mathrm{km\,s^{-1}}\), indicative of weak AGN activity, shocks, or outflows. For these candidates we infer modest warm-ionized outflow rates of \(1\!-\!4~\mathrm{M_\odot\,yr^{-1}}\) and kinetic powers \(\sim0.1\!-\!1\%\) of the AGN bolometric luminosity (from central \([\mathrm{O\,III}]\) or \(\mathrm{H}\alpha\)). These values place our sample at the low-energy tail of known AGN-driven outflows yet in continuity with \(\dot{M}_{\rm out}\)–\(L_{\rm AGN}\) scaling relations across \(0<z<6\). A completeness assessment shows MSA-3D is sensitive to AGN with \(L_{\rm AGN}\gtrsim10^{43}~\mathrm{erg\,s^{-1}}\), underscoring both the promise and current limitations of detecting weak AGN activity in distant galaxies with resolved spectroscopy.
\end{abstract}



\section{Introduction} \label{sec:Introduction}


Rest-frame optical nebular emission lines provide a powerful window into the physical conditions of the interstellar medium (ISM) in star-forming galaxies. The collisionally excited and recombination lines powered by young stars, active galactic nuclei (AGN), or shocks, trace the density, temperature, chemical, and ionization state of the ISM, and offer critical insights into star formation (SF) and feedback processes. The relative strengths of key emission lines, particularly when plotted as diagnostic line ratios, allow us to disentangle the dominant ionizing sources in galaxies and characterize their internal conditions.  
One of the most widely used diagnostic tools is the Baldwin–Phillips–Terlevich (BPT) diagram \citep{baldwin}, which uses [OIII] $\lambda$5007/H$\beta$ $vs.$ [NII]$\lambda$6584/H$\alpha$ (or [SII]$\lambda$6717,31/H$\alpha$) line ratios to distinguish between ionization from star formation, AGN activity, and low-ionization nuclear emission-line regions (LINERs).  In the local Universe, galaxies trace a well-defined star-forming sequence in this diagram, where decreasing [OIII]/H$\beta$ and increasing [NII]/H$\alpha$ reflect a progression of increasing metallicity and decreasing ionization parameter \citep{kewley19}, with AGN and LINERs occupying a distinct, separate region \citep{kewley01, kauffmann03}. 
For sources lacking complete sets of lines, alternative diagnostics such as the [NII]/H$\alpha$ versus EW(H$\alpha$) diagram, known as the WHAN diagnostic \citep{cidfernandes11}, are used to distinguish between star formation and AGN excitation source.

Over the past decade prior to JWST, ground-based near-infrared spectrographs and HST/WFC3 grism spectroscopy enabled optical emission line studies of galaxies at $z \sim 1$–3 \citep{steidel14, shapley15, coil15, momcheva16, sanders16, henry21}. These surveys revealed that high-redshift galaxies exhibit systematically lower metallicities, higher star formation rates (SFRs), harder ionizations and elevated gas velocity dispersions compared to local systems and fixed stellar mass \citep{shivaei15, schreiber19, wisnioski19}. Additionally, star forming galaxies at $  z > 1 $ consistently show offsets in the BPT diagram relative to their low-redshift counterparts, often with elevated [OIII]/H$\beta$ at fixed [NII]/H$\alpha$. This suggests contributions from one or more factors: harder ionizing spectra, evolving ISM conditions, and enhanced nitrogen enrichment at higher redshift \citep{steidel14, shapley15, sanders16,  masters16, shapley19, sanders23}.

With the launch of JWST, we have entered a new era of rest-optical diagnostics at high redshift. Thanks to the sensitivity and spectral coverage of JWST/NIRSpec and MIRI, rest-frame optical lines can now be studied out to $z \sim 9.5$. Early results from programs such as AURORA \citep{shapley25, sanders25, topping25}, CEERS \citep{sanders23, shapley23, backhaus24}, GLASS \citep{trump23}, JADES \citep{cameron23} and other campaigns \citep{nakajima23, curti23} confirmed continued offsets of high redshift galaxies ($z \sim 1 - 7$) in BPT space compared to the local sequence, and revealed a wide diversity in ionization conditions and metallicities, even among galaxies on the star-forming main sequence.

While these recent spectroscopic efforts have significantly advanced our understanding of the redshift evolution of ISM properties including excitation, ionization, and chemical abundances through integrated emission-line diagnostics, the lack of accompanying spatially resolved spectroscopy at high redshift has left the internal structure of galaxies largely unexplored. 
Integral field unit (IFU) spectroscopy is essential to dissect the local interplay between star formation, chemical enrichment, gas flows, and feedback, and to capture how nebular emission-line properties vary across different physical regions within individual galaxies at higher redshift. This has been demonstrated extensively at low redshift by several large IFU surveys such as MaNGA \citep{bundy15}, CALIFA \citep{sanchez16}, and PHANGS-MUSE \citep{emsellem22} and at high redshift by SINS using SINFONI IFU \citep[$z\sim 1-2; $][]{schreiber19, genzel11}. 
Those studies have revealed that galaxies often host spatially complex ionization structures, with the same galaxy simultaneously exhibiting AGN-like, LINER-like, and star-forming excitation in different regions depending on  local conditions \citep{sarzi10, belfiore16}. Resolved emission line maps are thus crucial to trace radial metallicity gradients, distinguish  ionization mechanisms between galaxy centers and outskirts, and detect weak AGN-driven outflow or shock signatures that may be unidentifiable in integrated global spectra.

Crucially, spatially resolved spectroscopy has been key in uncovering weak outflow or inflow signatures that are often missed in integrated light due to their low surface brightness or small velocity offsets \citep{roy18, roy21a, roy21b, venturi21, riffel24}. A striking example of this is MaNGA's discovery of ``red geyser'' galaxies where large scale biconical ionized gas outflows driven by weak AGNs and neutral gas inflows were only detected through spatial mapping of kinematics and ionization and remained hidden in SDSS single-fiber spectra \citep{cheung16, roy21a, roy21b, roy21c}.
This is now particularly timely in the high redshift regime. 
While JWST has revealed broad-line and heavily obscured AGNs at high $z$ \citep{harikane23, labbe23, maiolino24, kocevski25, ubler24, kokorev24, matthee24}, the incidence of AGNs that contribute only subdominantly to the galaxy’s emission remains unexplored, and can only be disentangled through spatially resolved spectroscopy.

In this paper, we present results from the \texttt{MSA-3D} program, which is a JWST/NIRSpec campaign using the slit-stepping mode of the Micro-Shutter Assembly (MSA) to acquire IFU-like spectroscopy.
Using extremely efficient multiplexed observations with 0.2$''$ MSA slitlets, this program provides high quality spatially resolved spectra for a large sample of 43 galaxies at $0.5 < z < 1.5$ simultaneously, in $<10\%$ of the time needed for an equivalent survey in the NIRSpec IFU mode \citep[see ][ for more details on the MSA-3D program]{barisic25, ju25}.
This offers the opportunity to spatially resolve the ionized ISM in typical star-forming galaxies at $z \sim 1$, reaching spatial scales of $<$1 kpc during the peak epoch of star formation.
Using high-resolution NIRSpec/MSA observations, we construct two-dimensional maps of key emission lines, mainly H$\alpha$, H$\beta$, [NII], [SII], and [OIII] across each galaxy, which enables a detailed investigation of internal gas morphology, ionization structure, and kinematics. We  find a range of ionized gas morphologies $-$ from smooth disks to patchy, clumpy distributions, and in some cases, spatial offsets between stellar continuum and ionized gas resolved maps. Such morphologies are consistent with those seen in clumpy, gas-rich galaxies at $z > 1$ \citep{elmegreen07, genzel11}, likely driven by disk instabilities, mergers, or local feedback.

Using slit-stepped NIRSpec cubes, we construct spatially resolved ionization diagnostic maps (BPT/WHAN) and construct averaged radial profiles of the principal nebular ratios  $-$ for e.g., [N II]/H$\alpha$, [O III]/H$\beta$, and [S II]/H$\alpha$, to trace chemical enrichment, ionization structure, and the physical state of the ionized ISM across each galaxy. 
We place these spatial trends in context by comparing their overall character with representative results from $z\sim0$ IFU surveys. This lets us track how gas excitation conditions vary in galaxies across redshifts without relying on one-to-one calibrations.

We further examine the link between local ionization conditions and ionized–gas kinematics by comparing spaxel–scale line ratios with the line-of-sight velocity and velocity dispersion ($\sigma$). We find a striking result: a spatially resolved correlation between [N II]/H$\alpha$ ratio and  $\sigma$ locally across sub-galactic regions in the entire sample, similar to trends noted in nearby systems \citep{dagostino19}. By this analysis, we also identify a small subset of systems that exhibit compact, high-excitation zones and elevated dispersions. On closer inspection, the spectra from these regions are inconsistent with pure star formation and instead point to mixed contributions from AGN photoionization, shock and outflow heating, and evolved (LINER-like) stellar sources, coincident with enhanced turbulence.  These spatially resolved diagnostics reveal weak AGN/outflow activity that would be difficult to recognize in integrated spectroscopy.  We measure associated mass‐outflow rates and kinetic energies, and derive the lowest–energy tail of AGN feedback signatures accessible with JWST via this slit stepping experiment.  This underscores the instrument’s power to expose small–scale feedback signatures and the internal dynamical complexity of $z>1$ galaxies.

This paper is organized as follows: Section \ref{sec:data} describes the NIRSpec MSA-3D observations, sample selection and emission line fitting procedures. Section \ref{sec:results} presents the main results, focusing on spatially resolved emission line maps, BPT diagnostics, and gas kinematics. In Section \ref{sec:discussion}, we discuss the implications of the results, and we summarize our conclusions in Section \ref{sec:conclusion}.

\section{Data} \label{sec:data}

\subsection{Observations and Sample selection}

This work is based on JWST NIRSpec spectroscopic observations from the Cycle 1 program GO 2136 (PI: T. Jones), which uses the NIRSpec Micro-Shutter Assembly (MSA) in slit-stepping mode to obtain spatially resolved spectroscopy of the sources. This approach enables the construction of pseudo-IFU datacubes, allowing us to spatially map rest-frame optical emission lines and  continuum across galaxy disks at kiloparsec scales. A detailed description of the slit-stepping strategy and the 3D datacube reconstruction pipeline is provided in \cite{barisic25}. 
The observations were taken using the G140H/F100LP grating and filter combination, providing high spectral resolution ($R \sim 2700$)  across a wavelength range of \(0.97\,\mu\mathrm{m} < \lambda < 1.82\,\mu\mathrm{m}\). This setup allows us to detect all the strong rest-optical nebular lines, primarily H$\alpha$, [NII]6584, [SII]6717, 6731, [OIII]5007, and H$\beta$ $-$ within a single grating for galaxies within $0.5 <z <1.7$. The reconstructed data cubes have a point spread function (PSF) that is approximated by a 2D Gaussian with FWHM = \(0.1^{\prime\prime} \times 0.2^{\prime\prime}\). These are sampled onto a spatial grid of 0.08$'' \times$ 0.08$''$, which corresponds to a spaxel size of $0.5-0.7$ kpc for the redshift range $0.5 < z < 1.7$.  


The parent sample consists of 43 star-forming galaxies at $0.5 < z < 1.7$, selected to ensure that multiple strong rest-optical lines fall within the observed wavelength window. The galaxies are located in the Extended Groth Strip (EGS)--a well-studied field with extensive multi-wavelength coverage. All targets benefit from existing photometric and spectroscopic data from HST, JWST, and Keck, due to their inclusion in major legacy surveys such as CANDELS \citep{koekemoer11}, 3D-HST \citep{momcheva16}, MOSDEF \citep{kriek15}, DEEP2 \citep{newman13}, and CEERS \citep{finkelstein24}. To ensure robust emission–line measurements across each galaxy, we visually inspected all data cubes and excluded five targets with inadequate signal-to-noise (S/N), specifically systems for which the brightest line (H$\alpha$ or [O\,\textsc{iii}]~$\lambda5007$) had $\mathrm{S/N}<2$ in more than $90\%$ of spaxels.
This results in a final sample of 38 galaxies, which forms the basis for all analyses in this work.

Figure.~\ref{fig:sfrms} illustrates the properties of the galaxy sample used in this work. We adopt  stellar mass and SFR estimates from the 3D-HST catalog \citep{momcheva16}, and use ancillary high-resolution imaging from JWST/NIRCam (CEERS) and HST/ACS+WFC3 (CANDELS) to contextualize the spatial distribution of line emission. We compute projected and deprojected galactocentric radii for each spaxel using structural parameters like position angle and axis ratio ($b/a$) derived from published catalogs based on GALFIT analyses of HST imaging \citep{vanderwel14}.

\subsection{Measuring emission line ratios and gradients} \label{subsec:fitting}

In this work, we analyze the spatially resolved ionization and kinematic properties of 38 star-forming galaxies at redshifts $0.5 < z < 1.7$, using JWST/NIRSpec MSA-3D spectroscopy. Our analysis focuses on extracting emission line fluxes, velocity fields, and velocity dispersion maps from the datacubes to construct physical diagnostics across each galaxy.

We model the strongest rest-frame optical emission lines, primarily [OIII]4959, 5007, H$\beta$, H$\alpha$, [NII]6548, 6584, and [SII]6717, 6731, by fitting single Gaussian profiles on a spaxel-by-spaxel basis. For each galaxy, we fit H$\alpha$, [NII], and [SII] lines simultaneously using a five-component Gaussian model plus a linear continuum. To reduce degeneracies in low S/N regions, we tie the central velocity ($v$) and velocity dispersion ($\sigma$) of all lines in every spaxel, and fix the flux ratio of the [NII]$\lambda$6584/6548 doublet to the theoretical value of 2.94. For galaxies where [SII] is not detected or redshifted out of the spectral range, we use a simplified three-component model for H$\alpha$ + [NII] alone.
A similar approach is adopted for the [OIII]$\lambda$4959,5007 and H$\beta$ complex, where we fit a three-component model with a shared $v$ and $\sigma$ across lines, and fix the [OIII]$\lambda$5007/4959 flux ratio to 2.98. All fits are performed independently in each spaxel. We include only those spaxels with signal-to-noise ratio $\mathrm{S/N}>3$ in all primary emission lines used -- specifically H$\alpha$, H$\beta$, [N\,\textsc{ii}]~$\lambda6584$, and [O\,\textsc{iii}]~$\lambda5007$. Spaxels failing this threshold in any of these lines are excluded from all 2D maps. 

To characterize the ionization conditions across galaxies, we employ two commonly used diagnostics: the BPT diagram \citep{baldwin, kewley01, kauffmann03} and the WHAN diagram \citep{cidfernandes11}. Where all four key lines--[OIII], H$\beta$, H$\alpha$, and [NII]--are available, we use the [NII]/H$\alpha$ vs. [OIII]/H$\beta$ BPT diagram to distinguish between HII region-like emission, composite excitation, LINER and AGN-dominated regions. In cases where H$\beta$ or [OIII] are not reliably detected, we instead use the WHAN diagram, which combines [NII]/H$\alpha$ with the equivalent width of H$\alpha$ to classify ionization sources. Together, these two diagnostics allow us to trace spatial variations in ionization mechanisms across each galaxy.

We also derive gas kinematics from the best-fit emission line profiles. The observed velocity dispersion ($\sigma_{\rm obs}$) from each Gaussian fit is corrected for instrumental broadening by subtracting the spectral resolution of the NIRSpec G140H grating in quadrature. We adopt a line spread function FWHM $\sim$ 5.2 \AA, appropriate for $R \sim 2700$, which corresponds to a velocity resolution of $\sigma_{\rm instrumental} \sim \rm 50-60 \ km \ s^{-1}$ across the observed wavelength range. The resulting $\sigma$ maps capture the intrinsic turbulent motion of ionized gas across each galaxy.

To quantify radial gradients in emission line ratios and velocity dispersion, we calculate the deprojected galactocentric distance of each spaxel from the galaxy center. Galaxy centers are defined as the peak of the continuum light, computed by collapsing the datacube along wavelength while masking strong emission lines. We adopt galaxy position angles and axis ratios (b/a) from available  HST-based  catalogs  \citep{vanderwel14}, allowing us to deproject each spaxel’s radial distance and construct radial profiles for individual line ratios and kinematic properties. Radial distances are then normalized by the galaxy’s effective radius ($\rm R/R_{\rm e}$) to compute radial gradients for each galaxy.

\section{Results} \label{sec:results}

\subsection{Global Properties}

\begin{figure*}
    \centering
    \includegraphics[width=\textwidth]{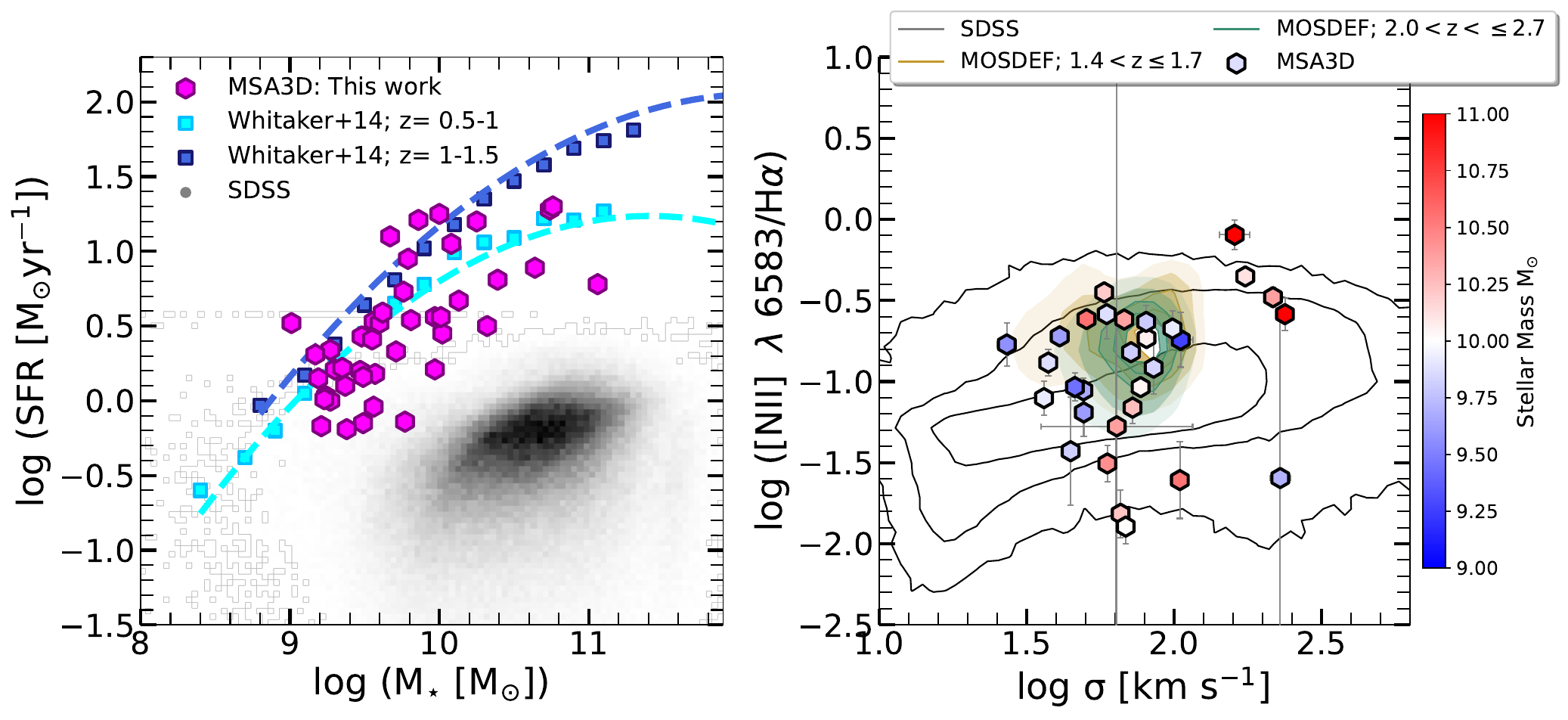}
    \caption{[Left] Star formation rate (SFR) versus stellar mass ($\rm M_{\star}$) for MSA-3D galaxies (magenta hexagons), derived via multi-wavelength SED fitting \citep{koekemoer11, stefanon17}. For comparison, we show the star-forming main sequence (SFMS) relations at $0.5 < z < 1$ (cyan) and $1 < z < 1.5$ (blue) from \cite{whitaker14}, along with the $z \sim 0$ SDSS galaxy population as grey 2D historgram. Our sample is consistent with the SFMS between z$\approx 0.5-1.5$. 
     [Right] Emission–line excitation, traced by $\log([\mathrm{N\,II}]\,\lambda6584/\mathrm{H}\alpha)$, (the horizontal axis of the WHAN diagram) $vs.$ gas velocity dispersion ($\sigma$), for MSA–3D galaxies. Both quantities are measured from integrated spectra co–added within $2R_e$ apertures. Only systems with [N\,\textsc{ii}] and H$\alpha$ detected within the observed spectral window are shown (colored circles). They are color coded by stellar mass.  Overplotted for reference are contours for SDSS galaxies at $z \approx 0$ (black) and MOSDEF galaxies at $1.4 < z < 1.7$ (yellow) and
     $2.0 < z < 2.7$  (green). They show almost no correlation in their integrated measurements. }
    \label{fig:sfrms}
\end{figure*}

We begin by characterizing the global properties of the MSA-3D sample. Understanding where these galaxies lie in relation to the star-forming main sequence is important, as both gas-phase metallicity and ionization conditions are known to correlate with stellar mass and star formation rate \citep{manucci10, sanders16, sanders21}. 
Figure.~\ref{fig:sfrms} (left panel) shows the distribution of our targets in the SFR–stellar mass plane. Stellar masses and star formation rates (SFRs) are taken from multiwavelength SED fitting to the available \emph{HST} photometry from the CANDELS survey, as reported in publicly released catalogs \citep{koekemoer11, stefanon17}. We adopt SFRs derived from SED fitting to ensure consistency across the sample, as the H$\alpha$ and H$\beta$  emission lines $-$ which are commonly used for dust-corrected line based SFR estimates $-$ are not often simultaneously observed within the wavelength coverage ($0.97-1.82 \mu$m) of the G140H/F100LP grating used in the MSA-3D observations. 

The MSA-3D galaxies (magenta hexagons), spanning $0.5 < z < 1.7$, exhibit stellar masses of $\rm log_{10} (M_\star/M_\odot) = 9-11$ and lie predominantly along the star-forming main sequence (SFMS) at $z \sim 1$. For reference, we include the parameterized main sequence relations from Whitaker et al. (2014) at $z = 0.5-1.0$ (cyan dashed line) and $z = 1.0-1.5$ (blue dashed line), along with the local SDSS relation (grayscale 2D histogram). Our sample shows closer agreement with the $0.5 < z < 1.0 $ SFMS relation with a mean offset of $\sim 0.27$ dex. In comparison, the average deviation of the MSA-3D SFRs from the   $ 1.0 < z < 1.5 $ relation of \cite{whitaker14} is larger at $\sim$0.5 dex, but still overall consistent except maybe a handful of sources where the offset goes to 1 dex. Our galaxies are thus representative of the typical star-forming population at $z \sim 1$, and are not extreme outliers in terms of star formation activity.

In the right panel of Figure.~\ref{fig:sfrms}, we examine the dependence of gas-phase excitation, traced by [NII]$\lambda$6584/H$\alpha$ (the x-axis of the traditional WHAN diagram) on the integrated gas kinematics  (velocity dispersion, $\sigma$) within 2 $\rm R_e$. Effective radii $\rm R_e$ are measured with \textsc{GALFIT} on available HST/F160W imaging, and verified to be consistent with the \cite{stefanon17} catalog. For each source we co-add all spaxels within  2$\rm R_e$ and fit the integrated spectrum with a Gaussian emission-line model plus a first-degree polynomial continuum.  Line centers and dispersions are tied across the primary lines (H$\alpha$, [NII]$\lambda \lambda$6548,84, [SII]$\lambda\lambda$6717, 32, [OIII]), following the per-spaxel fitting procedure described in Section 2. The resulting
[NII]/H$\alpha$ and velocity dispersion ($\sigma$) for all MSA-3D sources are shown as hexagons, colored by their stellar mass. The error bars indicate 1$\sigma$ uncertainties (16th–84th percentile ranges) in the reported values.

Across the MSA-3D sample we find no significant correlation between [NII]/H$\alpha$ and $\sigma$ (Pearson coefficient test: $r=0.27, \ p = 0.25$. For context, we overlay MOSDEF galaxies at $z \sim 1.5$ and $z \sim 2.5$ (yellow and green filled contours), and the SDSS locus (black contours), which likewise show weak or absent trends.  Coloring the MSA-3D points by stellar mass indicates that higher [NII]/H$\alpha$ preferentially occurs at higher $\rm M_{\star}$. This apparent positive trend is consistent with the mass–metallicity relation, given that [NII]/H$\alpha$ correlates with gas-phase metallicity, and mass and metallicity are correlated \citep{pettini04}. A secondary dependence of [NII]/H$\alpha$ on outflow-driven broadening in more massive systems may also contribute, resulting in enhanced $\sigma$ in a few sources, which we explore further in later sections. Overall, MSA-3D galaxies occupy the parameter space expected for main-sequence star-forming galaxies at $z\sim 1$, while extending to lower stellar masses than most previous spatially resolved surveys at similar redshifts.


\subsection{Integrated and Resolved BPT and WHAN diagrams at $z\sim1$}

\begin{figure*}
    \centering
    \includegraphics[width=\textwidth]{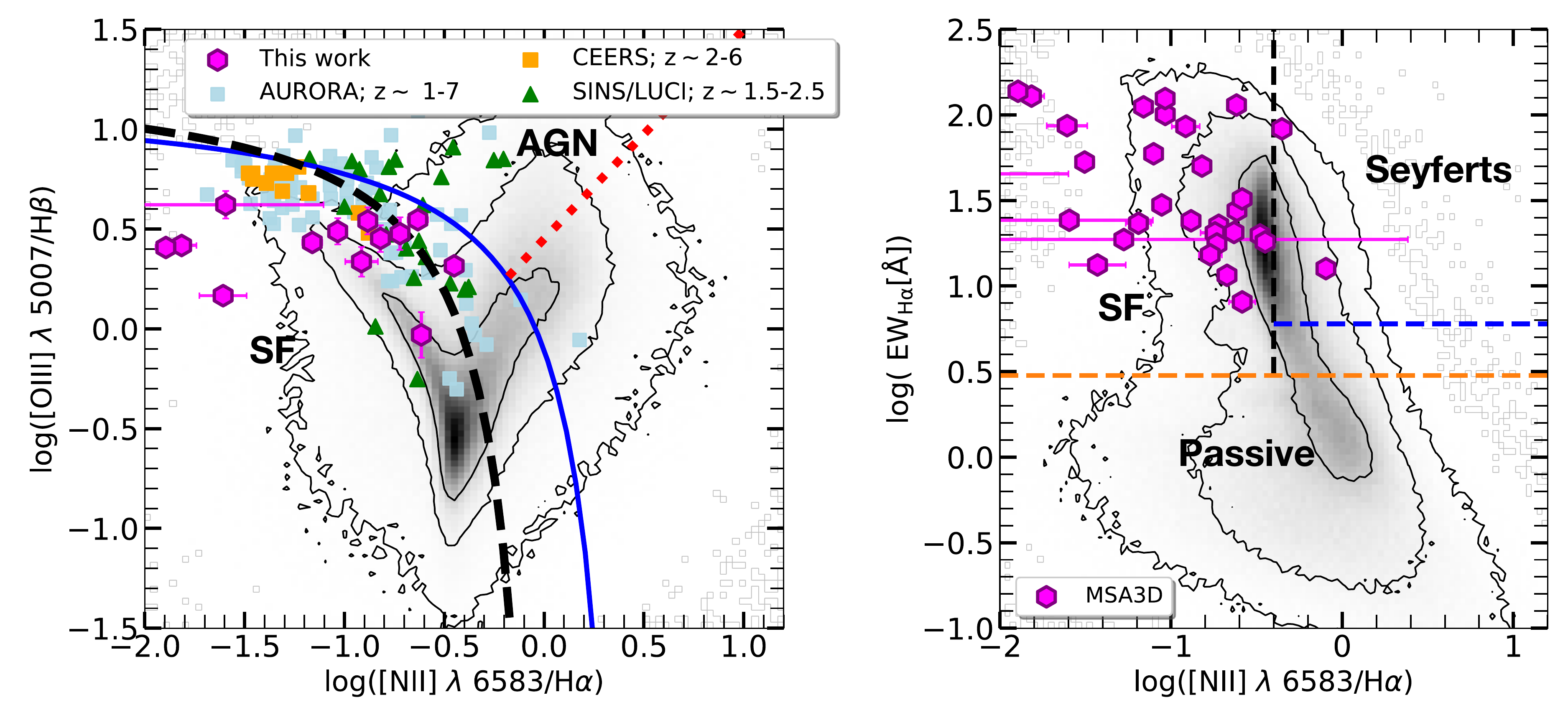}
    \caption{ [Left] Integrated BPT diagram for MSA-3D galaxies (magenta hexagons), where emission line fluxes are summed within 2$R_e$. 13 out of the 38 galaxies in our sample have wavelength coverage of all four emission lines required for this diagram.   For comparison, we overlay contours from SDSS (z $\sim$ 0), and include high-redshift star-forming samples from AURORA \citep[$z\sim1-7$;][]{shapley25}, CEERS \citep[$ z\sim2-6$;][]{sanders23}, and SINS/zC-SINF \citep[$z\sim 1.5 - 2.5$;][]{schreiber19}. Theoretical and empirical demarcation lines from \cite{kewley01, schawinski07, kauffmann03} are shown in blue, red and dashed black, respectively. MSA-3D galaxies fall below the AGN boundaries and are consistent with ionization predominantly by young stars in relatively low-metallicity environments from their integrated measurements. [Right] WHAN diagram \citep{cidfernandes11} showing log([NII]/H$\alpha$) $vs.$ H$\alpha$ Equivalent Width for a larger subset of the sample (31 sources) since G140H/100LP grating  captures $\rm H\alpha + [NII]$ for $\sim81\%$ of our sample. Nearly all of our galaxies lie well above the EW(H$\alpha$) = 3\AA \  threshold and within the star-forming region, with two sources (MSAID: 11843 and 13416) crossing the AGN/Seyfert classification boundary. }
    \label{fig:bpt_int}
\end{figure*}

\begin{figure*}
    \centering
    \includegraphics[width=\textwidth]{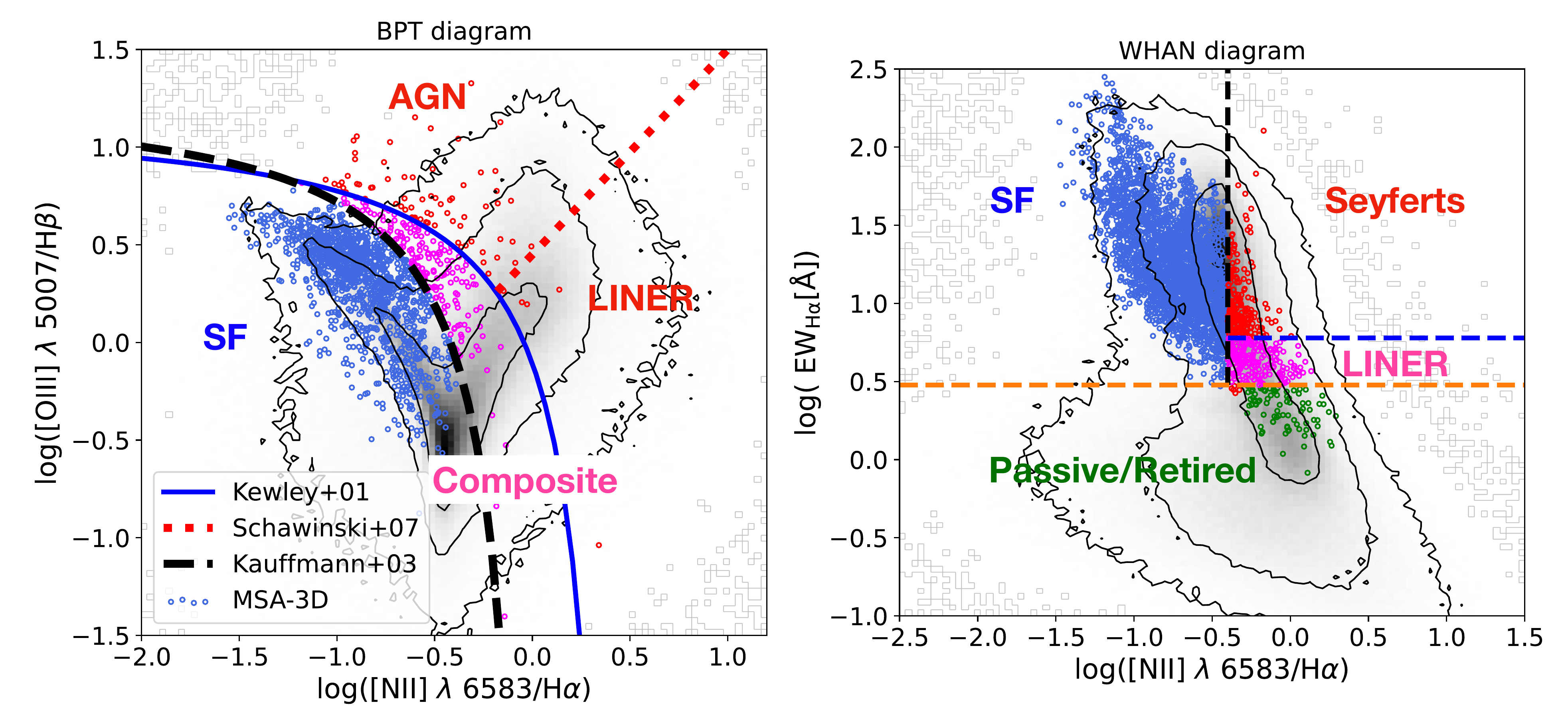} 
    \caption{Spatially resolved emission-line diagnostics for the MSA-3D sample. (Left) BPT diagram and (Right) WHAN diagram for individual spaxels for all MSA-3D galaxies. Displayed spaxels satisfy $\mathrm{S/N}>3$ for each of H$\alpha$, H$\beta$, [N\,\textsc{ii}]~$\lambda6584$, and [O\,\textsc{iii}]~$\lambda5007$, and any spaxels below this threshold in any line are masked. The flux measurements are obtained from line fitting, following the procedure described in \S\ref{subsec:fitting}. Spaxels are color-coded by their excitation mechanism based on the \cite{kewley01, kauffmann03} diagnostics: star formation or HII-region like in blue, AGN and Seyfert-like in red and composite ionization or LINER source in magenta. While the majority of spaxels occupy the star-forming locus, a non-negligible subset of spaxels extend into the composite/LINER and AGN-like regimes. This indicates that AGN, shocks, and mixed ionization sources can be missed in integrated spectra. Sensitive, spatially resolved spectroscopy is required to detect and disentangle these components. }
    \label{fig:bpt_resolved}
\end{figure*}

Building on the global properties discussed above, we now investigate the dominant ionization mechanisms within our sample using both integrated and spatially resolved emission line diagnostics. Specifically, we employ the classic [NII]/H$\alpha$ versus [OIII]/H$\beta$ BPT diagram \citep{baldwin, veilleux87}. Large spectroscopic surveys have shown that emission-line galaxies trace two distinct branches in the BPT diagram: a sequence at log([NII]/H$\alpha$) $<-0.3$  associated with star forming HII regions well reproduced by photoionization models of OB stars, and the branch at higher [NII]/H$\alpha$ linked to hard ionization sources like AGNs and Seyferts along with contribution from shocks and evolved stars from LINERs/LIERs \citep{kauffmann03, kewley01, stasinska06, belfiore16}. Thus, the BPT diagram using galaxy-integrated spectra demonstrates the dominant ionization mechanism responsible for ionizing the major portion of the ISM gas.

Figure.~\ref{fig:bpt_int} (left) presents the integrated BPT diagram for our sample (magenta hexagons), where line fluxes are summed within 2 times the effective radii ($R_e$). For comparison, we overlay contours from SDSS (z $\sim$ 0), and include high-redshift star-forming samples from AURORA \citep[$z\sim1-7$;][]{shapley25}, CEERS \citep[$ z\sim2-6$;][]{sanders23}, and SINS/zC-SINF \citep[$z\sim 1.5 - 2.5$;][]{schreiber19}. Theoretical and empirical demarcation lines from \cite{kewley01} and \cite{kauffmann03} are shown in blue and dashed black, respectively. 
Our galaxies fall below the \cite{kewley01} and \cite{kauffmann03} curves,  unlike some of the higher redshift sources from the AURORA survey, but display a broad range of [OIII]/H$\beta$ and [NII]/H$\alpha$ ratios consistent with photoionization from young stars. Over 85\% of the sample with simultaneous coverage of all four emission lines shows \(\log([\mathrm{O\,III}]/\mathrm{H}\beta) > 0.3\) and \(\log([\mathrm{N\,II}]/\mathrm{H}\alpha) < -0.8\)  in their integrated line ratios. This clear preference to lie in the upper left region in the BPT locus is in agreement with previous results at $z > 1$ \citep{shapley15, sanders16, strom17, curti23}, and reflects moderately low metallicities at higher redshift. Galaxies at $z > 1$, on average, have harder ionizing spectra, higher ionization parameters, and lower metallicity (corresponding to lower N/O abundance ratios) at fixed stellar mass compared to $z\sim 0$ galaxies from SDSS. Our targets are representative of the $z\sim1$ star-forming population in terms of excitation properties.

In order to include cases where [OIII] and/or H$\beta$ fall in the NIRSpec chip gap, remain undetected or lie outside the G140H/100LP spectral coverage, we complement the BPT analysis with the WHAN diagram \citep{cidfernandes11}, which only requires [NII] and H$\alpha$ emission lines and is widely used to classify sources based on ionization source and stellar population age. The integrated WHAN diagram (Figure.~\ref{fig:bpt_int}) is shown on the right panel, and has been derived for a larger subset of our sample due to the strategic choice of the G140H/100LP grating that includes H$\alpha$ + [NII] for 78\% of our sample. Nearly all galaxies in our sample lie above the EW(H$\alpha$) $= 3$\,\AA\ threshold commonly used to separate star-forming systems from LINER/AGN-dominated sources  \citep{cidfernandes11}. They  occupy the star-forming locus, indicating that their integrated emission-line spectra are governed by H\,\textsc{ii}-region excitation from young stellar populations, with little evidence for quasar-like AGN activity or post-AGB–dominated ionization. 
The two sources (MSAID: 11843 and 13416) with $\rm log ([NII] \lambda 6583/H\alpha) > -0.5$ fall in the ``AGN/Seyfert'' category and will be revisited in a later section.

Next, we construct spatially resolved BPT and WHAN diagrams using the line measurements described in \S\ref{subsec:fitting},  and shown in Figure.~\ref{fig:bpt_resolved}. All analyses use spaxels with $\mathrm{S/N}>3$ in the required emission lines, i.e.,  H$\alpha$, H$\beta$, [N\,\textsc{ii}]~$\lambda6584$, and [O\,\textsc{iii}]~$\lambda5007$. Individual spaxels are color-coded by classification: HII-like (blue), composite/LINER-like (magenta), and Seyfert-like (red) following the Kewley et al. (2006) diagnostic. 
While most spaxels cluster in the star-forming region, a non-negligible fraction ($\sim 11\%$) of  spaxels over the whole sample extends into the “composite/LINER” and AGN-like regions of the BPT diagram, particularly toward higher [OIII]/H$\beta$ at fixed [NII]/H$\alpha$. A similar trend is seen in the resolved WHAN diagram (Figure.~\ref{fig:bpt_resolved}, right), where most spaxels reside in the star-forming zone, but some extend toward $\rm log([NII]/H\alpha) > - 0.4$ and  $\rm EW_{H\alpha} > 3$\AA, indicative of transition regions or small, localized zones of shock/AGN or LINER-like excitation.
This diversity in spaxel-level excitation is consistent with trends seen in local IFU surveys like MaNGA and CALIFA, where galaxies classified as star-forming in integrated spectra often reveal central regions with low-luminosity AGN-like excitation and/or shock-driven ionization or emission from evolved stars near the outskirts  \citep{belfiore16, roy21a, wylezalek18}.
 While these zones typically occupy only a small area ($< 5 \rm  \ kpc^2$) of the galaxy, their detection in the resolved BPT and WHAN diagrams highlights the importance of spatial resolution: such weak, compact ionization sources are completely diluted in integrated light.

\subsection{Spatially resolved line emission} \label{subsec:clump}

\begin{figure*}
    \centering
    \includegraphics[width=\textwidth]{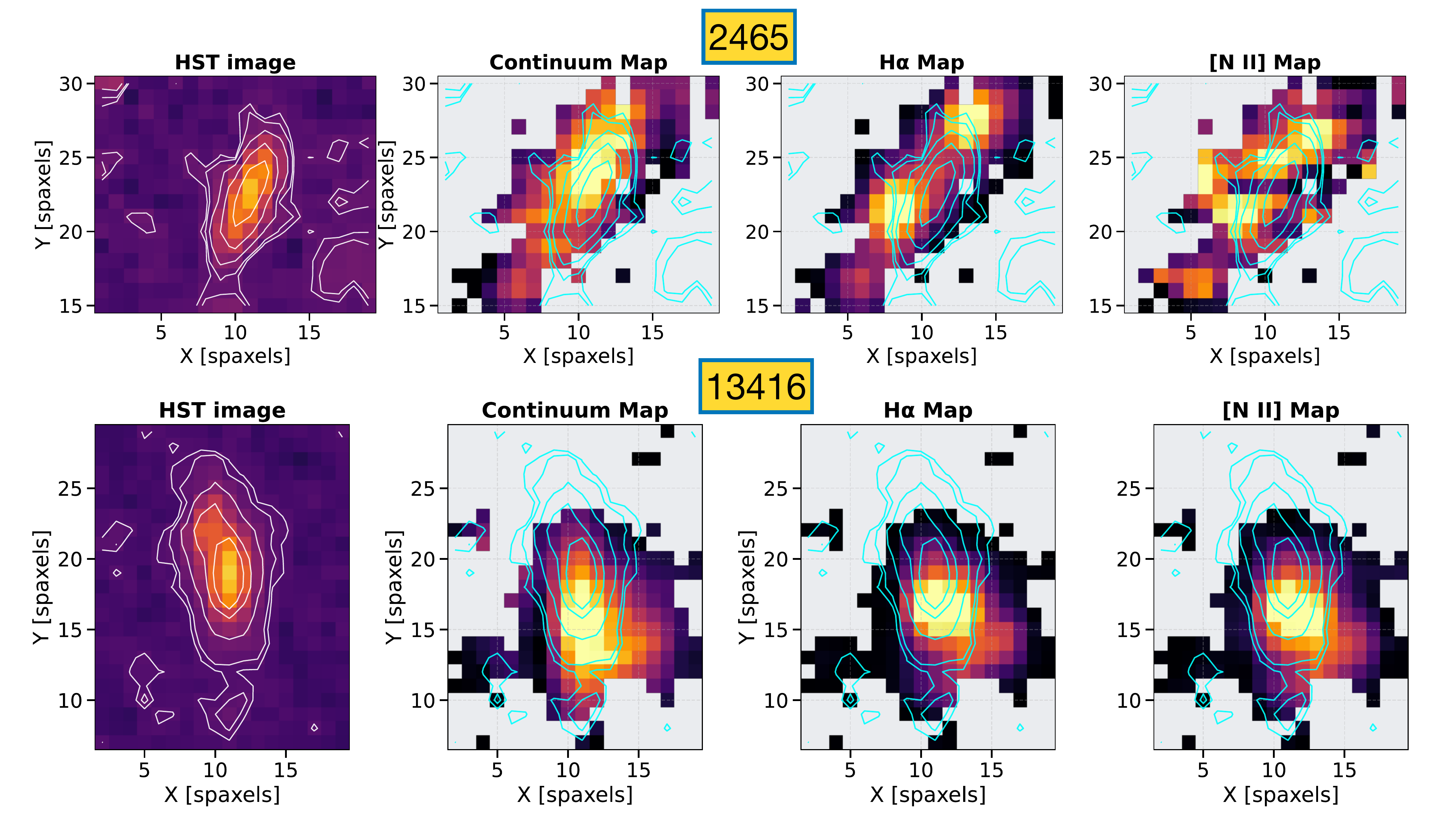}
    \caption{ Example two-dimensional maps for two representative galaxies (one in each row) from the MSA-3D sample, demonstrating the spatially resolved data quality achieved with the NIRSpec slit-stepping strategy. For each galaxy (top: ID 2465; bottom: ID 13416), the panels from left to right show: (1) the HST/WFC3 F160W image; (2) the rest-optical continuum map derived from the IFU datacube by masking strong emission lines and summing along the spectral axis; (3) the H$\alpha$ flux map; and (4) the [NII]$\lambda$6584 flux map. Emission–line fluxes are measured by Gaussian fits to individual spaxels that satisfy our S/N detection thresholds for all required lines, as detailed in \S\ref{subsec:fitting}. Overplotted contours trace the stellar morphology from the HST image, to enable direct spatial comparison. These examples highlight the morphological diversity in the ionized gas distribution across the MSA-3D sample. We encounter galaxies with well-ordered, disk-like H$\alpha$ emission aligned with the stellar distribution (similar to 2465), while a few sources show clumpy and offset ionized gas structures (13416). These reflect variations in gas excitation, kinematics, or dust geometry across the galaxy.}
    \label{fig:eg_map}
\end{figure*}

With JWST, we now obtain spatially resolved rest–frame optical spectroscopy for statistically meaningful samples of galaxies, enabling a robust dissection of their internal structure. From sub–kpc maps we can jointly trace ionized–gas excitation and kinematics, nebular abundances, and stellar continuum, yielding a physically coherent view of the gas and stars within each system.  
Figure.~\ref{fig:eg_map} shows example maps of two galaxies from our sample in the two rows. For each galaxy, we display the HST F160W image, continuum map, H$\alpha$ and [NII] flux maps where the S/N of each of the emission lines shown are greater than 3. These two example systems are representative of the diverse nature in the line emitting gas that are found in the entire sample. 
The top row presents a galaxy which shows a clumpy and irregular ionized gas distribution but the stellar continuum shows a disk-like centrally concentrated but elongated structure. However, the nebular emission lines exhibit knots and patches which are indicative of localized star-forming regions. Interestingly in this galaxy, the ionized gas shows two distinct peaks which are not spatially coincident with the peak of the stellar continuum.   It is possible that dust is obscuring parts of the ionized gas, making the distribution appear clumpy. About $\sim$40\% of the MSA-3D sample has clumpy, patchy ionized gas morphology resembling the galaxy at the top row. 

In contrast, the bottom row shows a galaxy with a much smoother and more centrally concentrated distribution in both the continuum and emission lines. Here, the gas and stars are largely spatially coincident, with emission from H$\alpha$, and [NII] following the same overall morphology as the stellar light. This suggests a more uniform star-forming disk, where ionized gas closely traces the stellar mass distribution. 
The diversity in gas morphology: from disturbed, clumpy gas distribution to smooth, disk-like structures,   provides important clues about the processes shaping star formation and feedback at z $\sim$ 1. To further understand how these variations manifest in the physical conditions of the gas, we now turn to the radial behavior of key emission line ratios across our sample.

\subsection{Radial Gradients in Emission Line Ratios} \label{radialbin}

\begin{figure}
    \centering
    \includegraphics[width=0.44\textwidth]{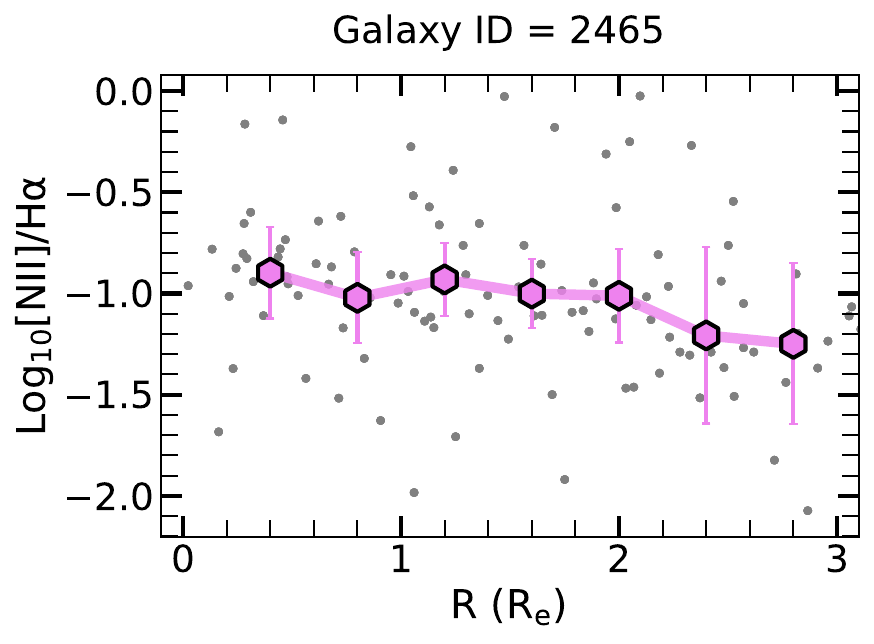}
    \caption{Radial profile of log([NII]/H$\alpha$) emission line ratio for a representative galaxy from our sample. We compute the deprojected radial distances for each spaxel and normalize by the effective radius ($R/R_e$). Gray points indicate the line ratio measurements for each individual spaxel, while magenta hexagons denote the mean values of the nebular ratio, computed within annular radial bins of equal width. We adopt this binned radial profile methodology, described in \S\ref{radialbin}, to characterize radial variations in ionization and abundance-sensitive line ratios. 
 }
    \label{fig:radial_eg}
\end{figure}

\begin{figure*}[!ht]
    \centering
    \includegraphics[width=\textwidth]{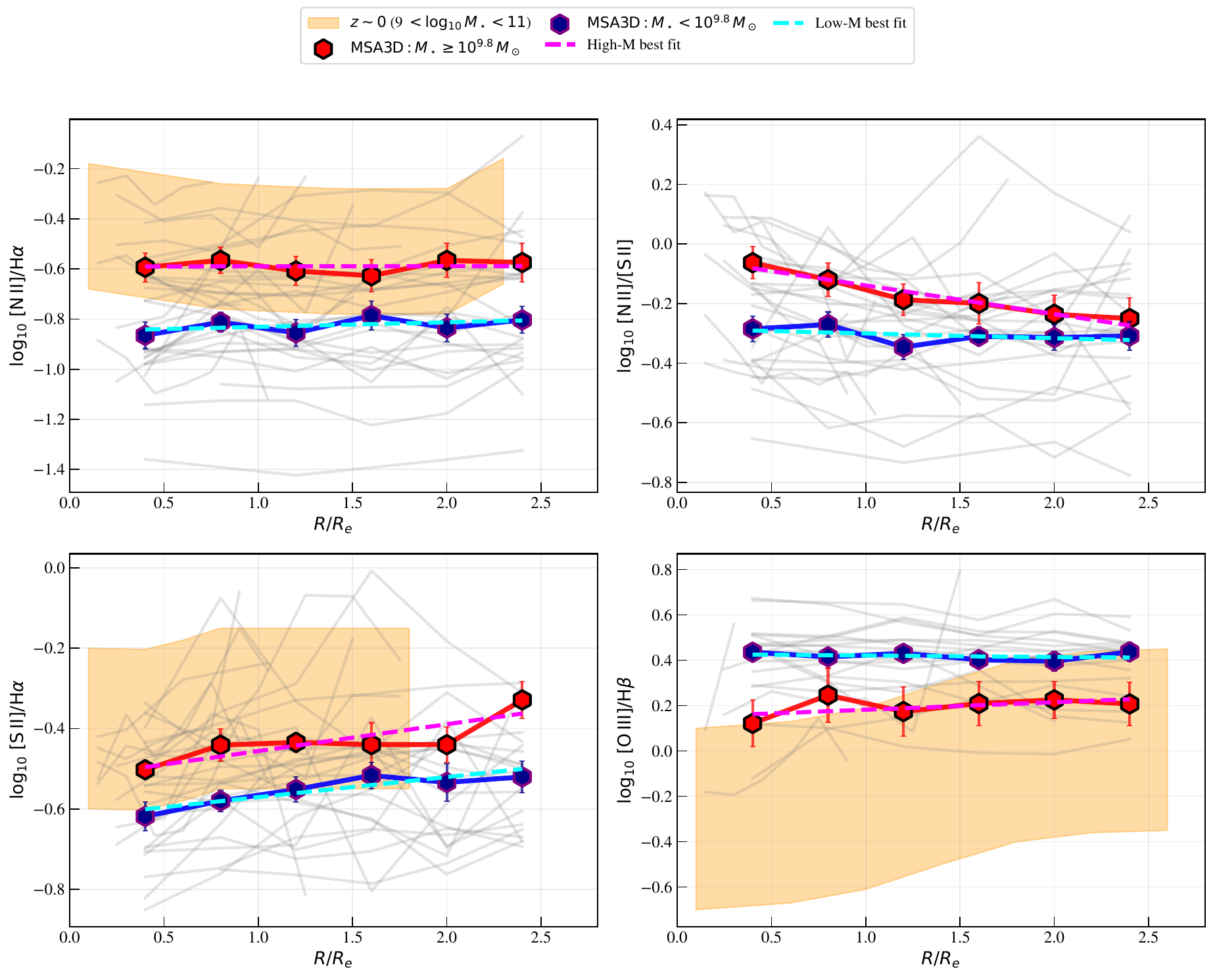}
    \caption{Binned radial profiles for four emission line ratios across the entire MSA-3D sample: $\rm log \ ([NII]/H\alpha)$ [top left],  $\rm log \ ([NII]/[SII])$ [top right], $\rm log \ ([SII]/H\alpha)$ [bottom left], and $\rm log \ ([OIII]/H\beta)$ [bottom right]. Individual galaxy binned radial trends, computed by the method described in \S\ref{radialbin}, are shown in gray. The blue and the red denote the median profiles for the high-mass ($\rm log \ M_{\star} \geq 9.8$) and low-mass ($\rm log \ M_{\star} < 9.8$) galaxies respectively. For comparison, the yellow shaded regions indicate the mean radial gradients with $1\sigma$ scatter observed in star-forming galaxies at $z\sim 0$ from the MaNGA survey \citep{belfiore16}. The MSA-3D galaxies exhibit systematically lower $\rm [NII]/H\alpha$ and $\rm [SII]/H\alpha$ ratios, and elevated $\rm [OIII]/H\beta$ compared to $z\sim0$ MaNGA star-forming galaxies, reflecting their lower metallicities and more extreme ionization conditions at $z\sim1$. The detailed description on the radial trends and their implications are discussed in the text. }
    \label{fig:radial}
\end{figure*}


Spatially resolved emission-line ratios provide powerful diagnostics of the physical conditions within the ISM, offering insight into gas-phase metallicity, ionization parameter, and the hardness of the ionizing radiation field. Radial gradients in key line ratios - such as [NII]/H$\alpha$, [SII]/H$\alpha$, [OIII]/H$\beta$, and [NII]/[SII] - probe the combination of the above quantities and trace the interplay between chemical enrichment, star formation, feedback processes, and gas accretion over cosmic time \citep{sanchez15, maiolino19}. 
These diagnostics capture how galaxy ISM properties evolve due to their past star formation history and ongoing baryon cycling within the disk.

For each galaxy in our sample of 38 MSA-3D sources, we compute the spaxel-by-spaxel line ratios [N\,\textsc{ii}]/H$\alpha$, [N\,\textsc{ii}]/[S\,\textsc{ii}], [S\,\textsc{ii}]/H$\alpha$, and [O\,\textsc{iii}]/H$\beta$. Here, we do not impose any S/N cut on any of the emission lines, to avoid biasing the line ratios toward only the brightest line–emitting regions.
 We compute de-projected radial distances for each spaxel and normalize by the effective radius ($R/R_e$). These per–spaxel measurements are then grouped into annular bins of equal radial width, and we construct average radial profiles by taking the mean of the spaxel–level line–ratio values within each annulus. This differs from an integrated profile, where one would first coadd the spectra within each annulus and then measure the line ratios on the resulting integrated spectrum. 
Figure.~\ref{fig:radial_eg} illustrates example radial profile of a representative galaxy from our sample. The gray points indicate individual spaxel measurements as a function of $R/R_e$, while the magenta markers show the mean values in the radial bins. We adopt this binned radial profile methodology throughout the paper to characterize radial variations in excitation and abundance-sensitive line ratios.
Figure.~\ref{fig:radial} presents the full set of binned radial profiles for the entire sample. Individual galaxy radial trends are shown in gray, while the median profiles for the high-mass ($\rm log \ M_{\star} \geq 9.8 \ M_{\odot} $) and low-mass ($\rm log \ M_{\star} < 9.8 \ M_{\odot}$) bins are over-plotted in red and blue, respectively. Error bars show the standard error of the mean  across galaxies in each radial bin.
 We also fit linear trends to the median profiles, shown in magenta and cyan for the high and low-mass bins respectively, to quantify the radial gradients of the median relations.
 For comparison, we overlay the mean radial gradients (with $1\sigma$ scatter) observed in star-forming galaxies at $z\sim 0$ from the MaNGA survey \citep[][; yellow shaded region]{belfiore16}. 



The [NII]/H$\alpha$ ratio serves as a widely used proxy for gas-phase metallicity, as the nitrogen-to-hydrogen abundance scales with oxygen abundance in metal-rich environments,  with some modest dependence on the ionziation parameter \citep{kewley02}. 
In our sample (Figure.~\ref{fig:radial}, top–left), the mass-averaged profile of $\rm log ([NII]/H\alpha)$ for the high-mass bin (red) is consistent with a flat gradient, with a best-fit slope consistent with zero (slope  $< 0.001 \pm 0.001 \ \rm dex \ Re^{-1}$). Individual systems (shown in gray) nevertheless show mildly negative trends. A dedicated metallicity-gradient analysis of the MSA3D sample by \cite{ju25} likewise finds mostly flat to negative gradients, with $\mathrm{d}Z/\mathrm{d}r$ spanning $-0.03$ to $+0.02~\mathrm{dex} \ \mathrm{kpc}^{-1}$, in agreement with our results.
A coherent picture emerges in which metallicity gradients steepen (become more negative) with increasing stellar mass, plausibly because more massive, disk–dominated systems retain and imprint inside–out enrichment, and this trend agrees with cosmological zoom–in simulations that include strong stellar feedback.
We refer the reader to \cite{ju25} for more detailed discussion on metallicity gradients. 
We do note that the  local galaxies ($z\sim 0$, shown in yellow) show a stronger negative gradient and higher absolute metallicities compared to our $z\sim 1$ systems, consistent with continued cosmic enrichment relative to our higher–redshift sample.

The [SII]/H$\alpha$ ratio, while sometimes used as a secondary metallicity indicator, is more sensitive to the ionization structure and gas phase conditions within the ISM, particularly in the partially ionized regions. This ratio is especially effective in distinguishing emission from classical HII regions and low-ionization (LI(N)ER-like) excitation. Our sample shows a mean log([SII]/H$\alpha$) $\sim$ -0.5 for the more massive galaxies, and $\sim$ -0.6 for lower mass systems. These values are broadly consistent with ionization from young stars in typical star-forming regions. For comparison,  \cite{belfiore16} reported a mean  value of log([S II]/H$\alpha$) $\sim -0.4$ for star forming regions in $z\sim 0$ galaxies, with a clear separation from LINER-like systems, which exhibit elevated values of log([S II]/H$\alpha$) $\sim -0.1$. The nebular emission in our sample is thus consistent with star formation driven ionization. 

log ([SII]/H$\alpha$) ratio shows a clear positive gradient, with best-fit slope ranging between $(0.050 - 0.067) \pm 0.016 \ \rm dex \ Re^{-1}$. This behavior contrasts with the nearly flat or midly declining [NII]/H$\alpha$ gradients and suggest radial changes in ionization state rather than metallicity alone. A similarly positive radial profile  has been observed in local galaxies  \citep{belfiore16, zhang17}, and is often interpreted as evidence for an increasing contribution from diffuse ionized gas (DIG) at larger galactocentric distances.
DIG-dominated regions enhance low-ionization forbidden lines relative to recombination lines like H$\alpha$, particularly in low surface brightness environments, thus elevating [SII]/H$\alpha$ with radius \citep{sanders17}. 
The observed radial profile supports a scenario where classical HII regions dominate the ionization in the central regions, while more extended, DIG emission  increasingly contributes in the outskirts \citep{zhang17, lacerda18}.
We note that shocks and low-level outflows can further boost [SII]/H$\alpha$, and we present such examples on a case by case basis on later sections.

Figure.~\ref{fig:radial} (top-right) shows a [NII]/[SII] shows a negative gradient, with slope $\rm = -0.096 \pm 0.013 \rm \ dex \ Re^{-1}$ in the high mass bin to $\rm = -0.016 \pm 0.011 \rm \ dex \ Re^{-1}$) in the low mass bin. A declining [NII]/[SII] points to a declining N/S, and, by extension, N/O abundance ratio with radius in the high–mass bin. Since Sulphur is an $\alpha$ element that broadly tracks oxygen, and nitrogen acquires a strong secondary component at higher metallicity (delayed AGB enrichment), log ([NII]/[SII]) serves as a useful proxy for N/O. 
 The steep negative trend therefore points to centrally enhanced nitrogen enrichment relative to $\alpha$ elements, consistent with inside–out chemical evolution and longer chemical processing times in the inner regions. Placed along side the nearly flat $\rm [NII]/H\alpha$ ratios, this implies that the overall metallicity (O/H) varies only weakly with radius, on average, but the relative abundance pattern varies. The galaxy centers are comparatively N-rich. Secondary effects like diffuse ionized gas (DIG) or presence of shocks may further diminish the [NII]/[SII] at larger radii, but the dominant mass dependent gradients are more naturally explained by the radial N/O gradient. 
Our interpretation is consistent with local benchmarks. \cite{berg20} report similarly negative radial gradients in N/O for nearby galaxies and attribute them to the delayed, largely secondary production of nitrogen in intermediate–mass stars, in contrast to the prompt $\alpha$–element yields (O and S) from core–collapse supernovae. In an inside–out framework, this time–delay naturally produces centrally enhanced N/O and a declining [NII]/[SII] with radius, in line with our measured slope. We note, however, that [NII]/[SII] can also respond to changing ionization structure because the lines arise from different zones, and thus variations in ionization parameter or diffuse ionized gas can modulate the ratio beyond pure abundance effects \citep{sanders23}.

Figure.~\ref{fig:radial} (bottom right panel) shows the [OIII]/H$\beta$ ratio, which is another diagnostic that depends on  ionization parameter, hardness of the radiation field and also gas phase metallicity. Lower metallicity environments typically host hotter, younger stellar populations with harder ionizing spectra and less efficient cooling, which enhance [OIII] emission relative to Balmer recombination lines. 
Our sample exhibits a clear mass dependence: low-mass galaxies show elevated [OIII]/H$\beta$ ratios with mean log ([OIII]/H$\beta$) $\sim 0.4$, and higher-mass galaxies lie systematically lower at a mean of 0.1. This trend is consistent with previous findings that low-mass galaxies at high redshift are both less chemically evolved and tend to host more extreme ionization conditions, which is further supported by the lower [N II]/H$\alpha$ in the BPT diagram \citep{sanders21, curti23}. 

The radial profiles of [OIII]/H$\beta$ in our sample are generally flat to mildly rising, with a best fit slope of $\rm \rm 0.0329 \pm  0.0228 \ dex/Re$ for the high mass bin.  Although only marginally positive, this behavior is expected in systems with flat or mildly negative  metallicity gradients, as lower metallicity reduces cooling, raises the electron temperature, and boosts the [OIII]/H$\beta$ ratio at larger R/$R_e$. So this is consistent with the O/H gradients discussed by \cite{ju25}.  Comparable flat or weakly positive [OIII]/H$\beta$ gradients have been reported in both local galaxies and high redshift samples \citep{belfiore17, sanders23, claeyssens23, kashino17}.
A more comprehensive analysis of gas-phase metallicity gradients and their stellar mass dependence in the MSA-3D sample is presented in \cite{ju25}.
We note that [OIII]/H$\beta$ also responds to the ionization parameter ($U$) and spectral hardness. A systematic decrease of $U$ with radius would tend to lower [OIII]/H$\beta$, while harder radiation fields or leakage can counteract this effect.  Altogether, these results indicate that the processes shaping metal production, transport, and ionization structure are already well established by $z\sim 1$, when they are still assembling their stellar mass and metal reservoirs.

\subsection{Kpc-scale spatially resolved correlation between [NII]/H$\alpha$ and velocity dispersion}

Radial gradients of emission line ratios, discussed in the previous section, provide important constraints on how gas-phase metallicity, excitation, and ISM conditions evolve with galactocentric distance. However, spatially resolved, spaxel-by-spaxel diagnostics can uncover more localized physical processes. Here, we investigate the spatially resolved connection between the nebular line ratios -- [NII]/H$\alpha$ and [OIII]/H$\beta$, that are sensitive to both metallicity and ionization, and the ionized gas velocity dispersion ($\sigma$), which reflects the dynamical state of the gas and can be elevated by turbulence, shocks, or feedback-driven motions.

\begin{figure*}
    \centering
    \includegraphics[width=1\textwidth]{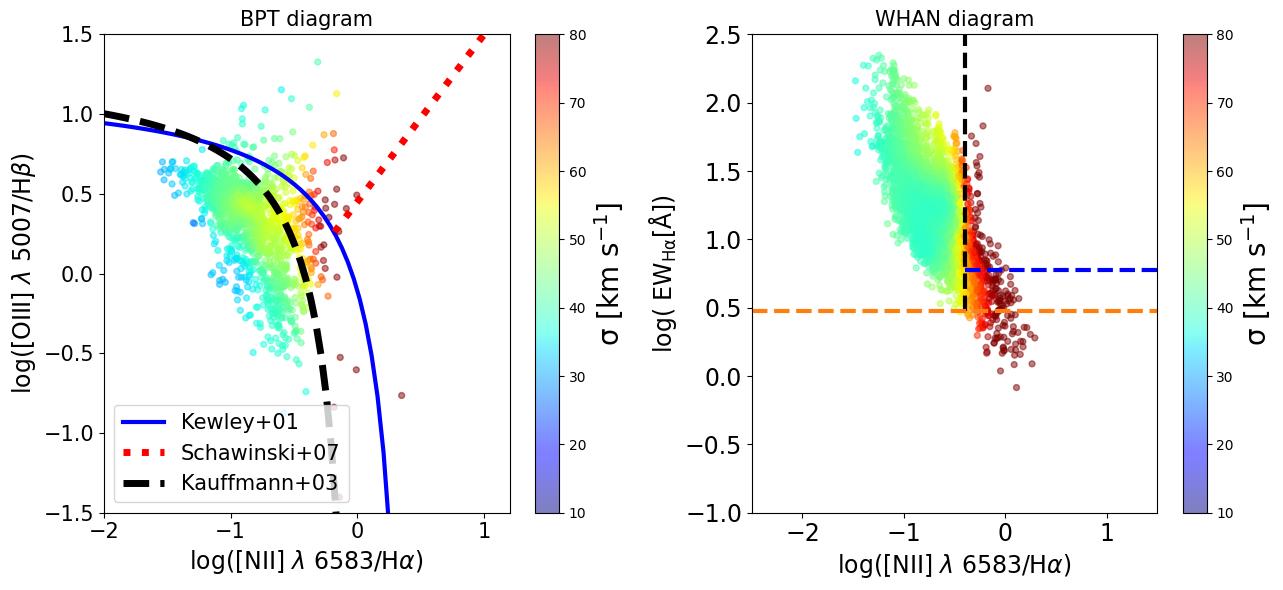} 
    \caption{ Spatially resolved BPT (left) and WHAN (right) diagrams color-coded by velocity dispersion ($\sigma$) for all spaxels in the MSA-3D sample with S/N of the required emission lines $>$ 3. We observe a clear increase of [NII]/H$\alpha$ with $\sigma$, consistent with enhanced excitation in high-dispersion regions. Elevated $\sigma$ is often associated with shocks, outflows, or AGN photoionization, and they preferentially lie in the Seyfert/AGN loci in both diagrams. This spaxel-by-spaxel view reveals weak, locally confined (low surface brightness) non-stellar excitation embedded within otherwise star-forming galaxies.}
    \label{fig:bpt_whan_sigma}
\end{figure*}

Figure~\ref{fig:bpt_whan_sigma} shows the distribution of individual spaxels from the full MSA-3D sample in the BPT-$\sigma$  and WHAN-$\sigma$  planes. We include every spaxel with $\mathrm{S/N}>3$ for each of H$\alpha$, H$\beta$, [N\,\textsc{ii}]~$\lambda6584$, and [O\,\textsc{iii}]~$\lambda5007$, irrespective of its galactocentric location. A clear positive correlation emerges: higher velocity dispersion coincides with elevated [NII]/H$\alpha$ in the WHAN diagram, while low-$\sigma$ regions tend to show lower line ratios. Part of this trend likely reflects the resolved mass-metallicity relation seen at both low and high redshift \citep{rosales12, sanchez13}, where regions of higher stellar surface density host more metal-enriched gas. Since stellar mass surface density correlates with kinematics \citep{cappellari06}, high-$\sigma$ regions naturally map to higher metallicity and thus higher [NII]/H$\alpha$.

However, deviations from this baseline can arise where shocks, outflows, or low-luminosity AGN locally boost both $\sigma$ and the excitation state. In particular, spaxels with elevated $\sigma$ that fall in Seyfert/AGN loci of the BPT/WHAN diagnostics likely trace compact, energetically active zones embedded within otherwise star-forming disks. Indeed, we see spaxels classified as AGN from the BPT/WHAN typically exhibit higher $\sigma$. These spatially resolved diagnostics therefore provide a sensitive means to uncover weak, low surface brightness non-stellar excitation that would be diluted in integrated spectra. A comprehensive analysis of emission-line kinematics for the full sample will be presented in a follow-up paper (Espejo Salcedo et al. in prep.). Here we identify sources hosting spaxels with high $\sigma$, enhanced $\rm log \ ([NII]/H\alpha)$, and lying in the AGN/Seyfert classifications in the BPT or WHAN diagrams, and discuss them on a case-by-case basis below.

\subsection{Case studies: Candidates with possible weak AGN activity}

We demonstrate in Figure.~\ref{fig:bpt_int} that the MSA-3D sample predominantly includes typical star forming galaxies with young, OB stars from HII regions largely contributing to the ionization of the ISM. Only two galaxies with ID 11843 and 13416 are consistent with AGN/LINER-like excitation from the integrated WHAN/BPT ($\rm log([NII]/H\alpha) > -0.4$) in the 2$\rm R_e$ aperture-summed spectrum. These two galaxies stand as candidates featuring ionization signatures of either low luminosity AGN activity  or shocked gas from centrally driven gas outflows.

On the other hand, Figure.~\ref{fig:bpt_resolved} demonstrates that if we deconstruct every galaxy and compute nebular line ratios in each spatially resolved element/spaxels for the whole sample, a bigger 
subset of galaxies may reveal localized signposts of weak AGN activity or shock-heated signatures which otherwise remain blended in their integrated spectra. Hence, we scan every galaxy and identify localized spaxel regions (3 contiguous spaxels in a box or $0.24'' \times 0.24''$ box at the minimum to ensure statistical significance) with Seyfert/LINER-like excitation accompanied by enhanced ($\geq 80 \ \rm km \ s^{-1}$) gas velocity dispersion. 
We also compute the observed second velocity moment of the ionized gas,
$ \rm V_{RMS} = \sqrt{v^2 + \sigma^2} \ km \ s^{-1}$,
where v is the line-of-sight velocity offset from the systemic redshift
measured from the H$\alpha$ fits, and $\sigma$ is the Gaussian velocity dispersion.
Sources with $\mathrm{V_{RMS}} \gtrsim 250~\mathrm{km\,s^{-1}}$ exhibit kinematics that are
too large to be explained by gravitational motions alone over the stellar-mass
range considered here, and are therefore strongly indicative of outflowing,
non-circular gas motions. We impose $\mathrm{S/N}>3$ in all constituent lines (including [N\,\textsc{ii}] and H$\beta$) to ensure that our line–ratio and $\sigma$ measurements and their uncertainties are reasonable. Errors in line ratios are computed by standard error propagation of the individual line flux errors.
Following these criteria, we find 6 such AGN or outflow or LINER candidates. In this section, we present those six cases in detail along with a pure star forming galaxy as a comparison source.

\begin{figure*}
    \centering
    \includegraphics[width=\textwidth]{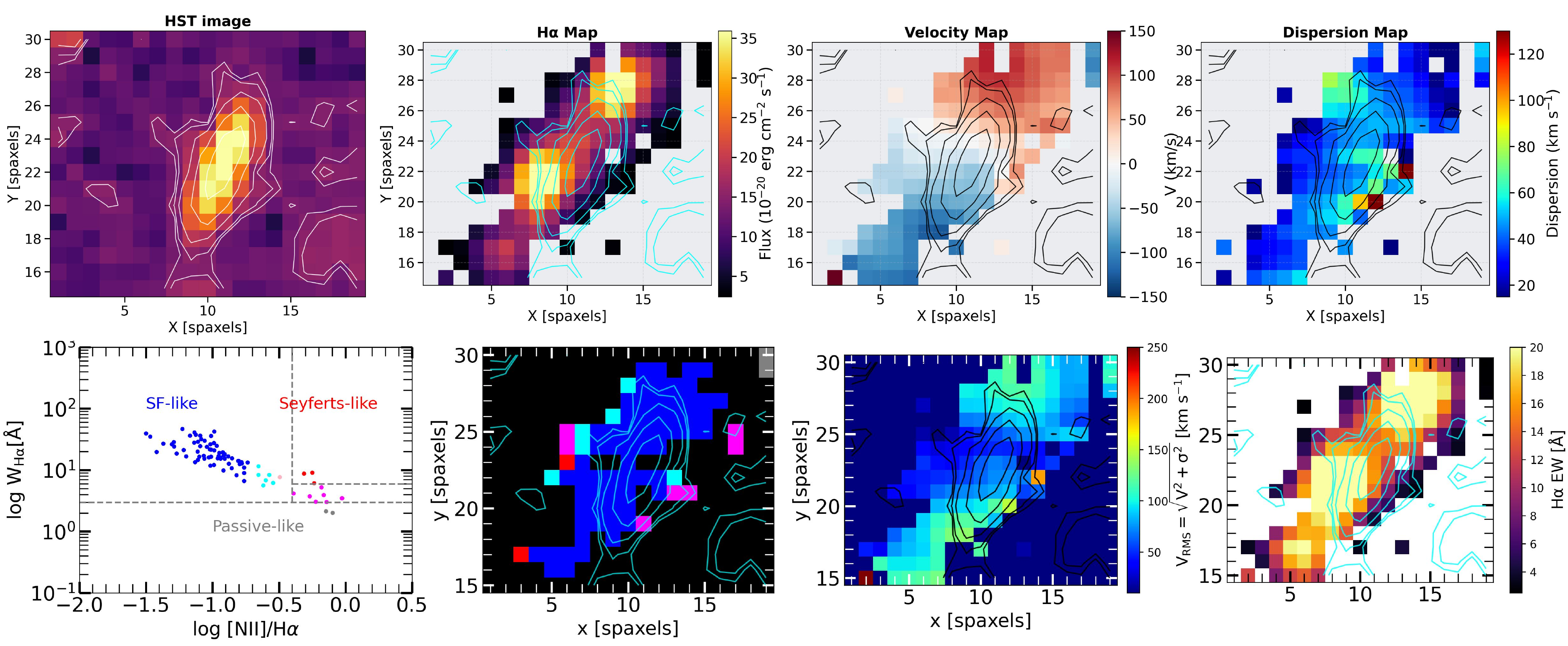}
    \caption{Detailed maps and spatial diagnostics of a purely star forming galaxy (ID: 2465). 
 Top row (left to right): HST F160W image, H$\alpha$ flux map, ionized gas (H$\alpha$) velocity field, and velocity dispersion map derived from Gaussian fits to H$\alpha$ emission line. Cyan and black contours trace stellar  isophotes from the HST image.  Bottom row: [Left] WHAN diagram showing individual $\rm S/N > 3$ spaxels color-coded by ionization class: star-forming (blue), Seyfert-like (red), LINER-like (magenta) and passive (gray), based on [NII]/H$\alpha$ ratio and H$\alpha$ equivalent width. Cyan and pink colors represent intermediate/ transition zones bridging the purely star formation regions to AGNs/LINERs. [Second from left] The distribution of the color coded spaxels projected on the galaxy. [Third from left] 2D map of the second velocity moment: $\rm V_{RMS} = \sqrt{v^2 + \sigma^2}$, which represents the dynamic motion of the gas. [Right] H$\alpha$ equivalent width (EW) map. Over 92\% of spaxels in this galaxy fall into the blue or cyan categories which suggests that the ionization is dominated by photoionization from star-forming regions. The ionized gas velocity field looks smooth and symmetric with $\rm V_{RMS} < 100 \ km \ s^{-1}$. Hence the gas kinematics is consistent with ordered rotation in a purely star forming disk galaxy.  }
    \label{fig:2465_map}
\end{figure*}

\subsubsection{ID: 2465 -- a purely star forming galaxy}

Figure~\ref{fig:2465_map} presents spatially resolved diagnostics of  a representative example of a star-forming system (ID: \texttt{2465}). The top row displays the HST F160W imaging and resolved ionized gas properties $-$ the H$\alpha$ flux, line-of-sight velocity field, and velocity dispersion map obtained from gaussian fits to the H$\alpha$ line (see \S\ref{subsec:fitting} for details). The bottom row shows: (i) the WHAN diagram constructed from all spaxels, (ii) the spatial location of those spaxels in the galaxy, color-coded by their WHAN classification, (iii) the second velocity moment ($\rm V_{RMS} = \sqrt{v^2 + \sigma^2}$), and (iv) the H$\alpha$ equivalent width (EW) map.
Spaxels are color-coded and categorized according to their position in the WHAN diagram based on log([NII]/H$\alpha$) and H$\alpha$ EW: 

\begin{itemize}[itemsep=-1.5pt, parsep=1.5pt]
\item log([NII]/H$\alpha$) $< -0.7$ and $\rm EW_{H\alpha} > 3$\AA: blue (pure star-forming)
\item $-0.7 \leq$ log([NII]/H$\alpha$) $< -0.5$ and $\rm EW_{H\alpha} > 3$\AA: cyan (SF-intermediate)
\item $-0.5 \leq$ log([NII]/H$\alpha$) $< -0.4$ and $\rm EW_{H\alpha} > 3$\AA: pink (transition zone)
\item log([NII]/H$\alpha$) $\geq -0.4$ and $3 < \rm EW_{H\alpha} < 6$\AA: magenta (LI(N)ER-like)
\item log([NII]/H$\alpha$) $\geq -0.4$ and $\rm EW_{H\alpha} > 6$\AA: red (AGN/shock-dominated)
\end{itemize}

We interpret the blue spaxels as tracing classical HII regions ionized by young, bright OB stars, while red spaxels are likely associated with AGN or fast radiative shocks. Magenta spaxels may reflect LI(N)ER-like excitation from evolved stellar populations (e.g., post-AGB stars) or slow moving shocks. Cyan and pink regions represent intermediate regimes transitioning from star formation to harder ionizing sources.

In the case of galaxy 2465, over 92\% of spaxels fall into the blue or cyan categories which suggests that the ionization is dominated by photoionization from star-forming regions. The few pink/magenta/red spaxels do not form coherent structures ($\leq$ 3 contiguous spaxels) and fail to meet our criteria for physical significance. Those may arise from noise or fitting artifacts. 
The gas kinematics give additional clues regarding the nature of this system. The ionized gas velocity field looks smooth and symmetric, consistent with ordered rotation. The kinematic major axis from the H$\alpha$ velocity field is aligned with the stellar disk seen in the HST image. The measured velocity dispersion ($\rm \sigma \leq 65 \ km \ s^{-1}$ and $\rm V_{RMS} \leq 130 \ km \ s^{-1}$) is modest, further supporting a dynamically settled disk in which ordered rotation dominates over disordered motion. We therefore classify this system as a purely star-forming galaxy with no evidence for AGN activity or kinematically disturbed gas from outflows/shock heated gas.

\begin{figure*}
    \centering
    \includegraphics[width=\textwidth]{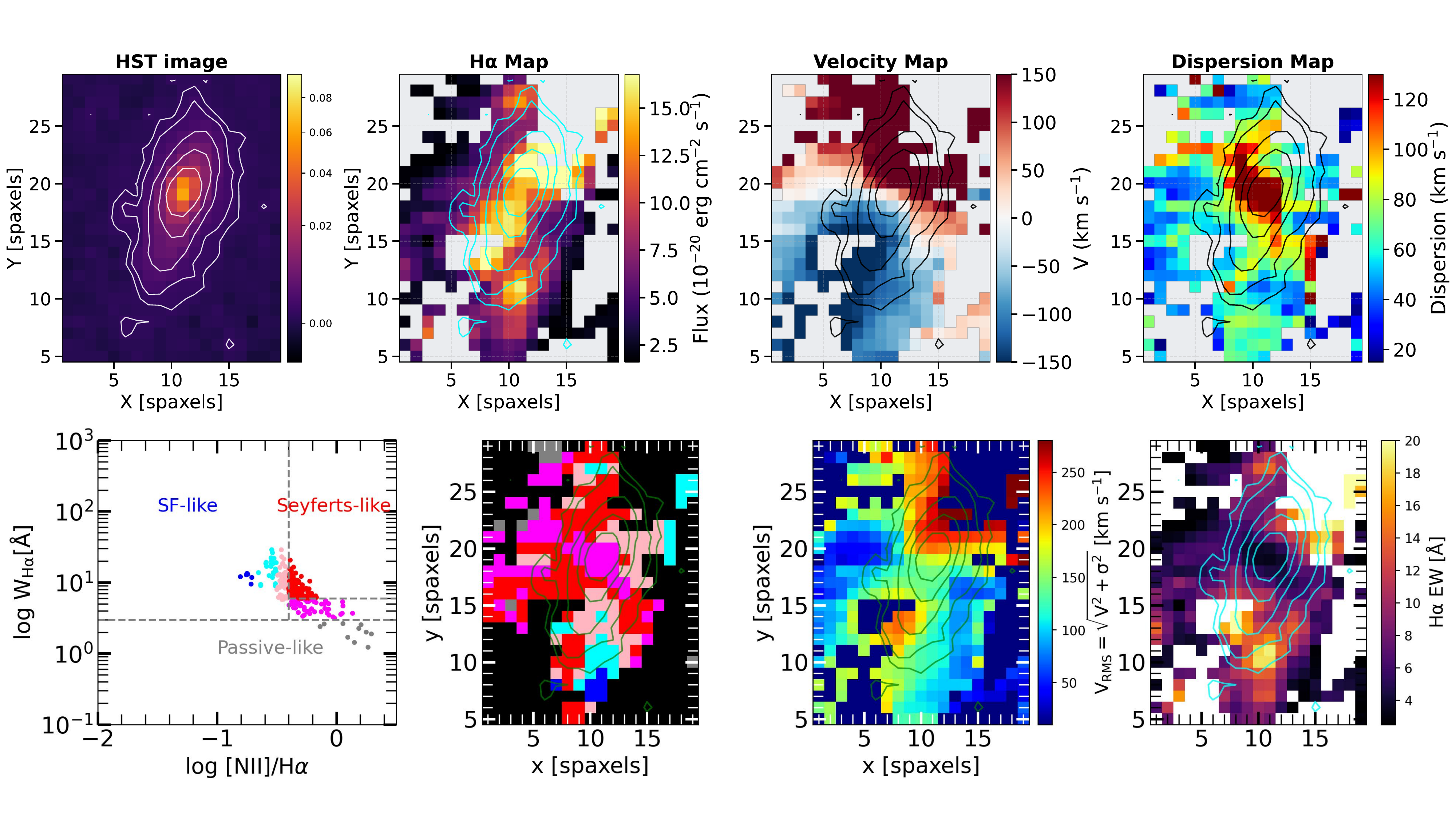}
    \caption{Spatially resolved maps for galaxy 9960. The panel layout and diagnostics shown are identical to those in Figure.~\ref{fig:2465_map}. The HST image suggests an inclined disk, but the velocity and dispersion maps reveal strongly disturbed ionized gas, with high-$\sigma$ regions outlining a roughly biconical structure lying parpendicular to the disk. WHAN-classified spaxels show a concentration of AGN/LINER-like excitation that is roughly co-spatial with elevated velocity dispersion. This object hosts turbulent outflow driven by the central engine rather than a purely rotation-supported disk. }
    \label{fig:9960_map}
\end{figure*}

\subsubsection{ID: 9960 \& 6848 -- low luminosity AGNs driving outflows}

We now present spatially resolved emission line diagnostics for candidates which exhibit signatures of AGN activity and disturbed ionized gas kinematics due to outflows. Figure~\ref{fig:9960_map} displays the maps for galaxy ID \texttt{9960}, where the panels and the symbols are as in Figure.~\ref{fig:2465_map}.  The HST imaging suggests an inclined disk, yet the line-of-sight kinematics—peak velocities ($v \geq 150 \ \rm km \ s^{-1}$) and dispersions
($\sigma \geq 120 \rm \ km \ s^{-1}$), indicate a dynamically disturbed,
highly energetic system rather than a smooth rotating disk. Although a BPT diagram cannot be constructed because
[OIII]$\lambda$5007 falls within a detector chip gap, we instead
employ the WHAN diagnostic.

The spatial distribution of WHAN-classified spaxels reveals a central concentration of AGN/shock-like ionization (red), transitioning radially outward through magenta and pink zones indicative of LI(N)ER-like or mixed excitation from evolved stars and shocks. This suggests a gradient from AGN photoionization to shock-dominated regimes, with possible contribution from evolved stars towards the outskirts $-$ similar to LINER/LIER emission seen in $z\sim 0$ galaxies \citep{belfiore16, roy18, roy21a}. The excitation structure is extended and asymmetric, and the H$\alpha$ equivalent width map appears clumpy, consistent with patchy, non-uniform ionizing sources.

\begin{figure}
    \centering
    \includegraphics[width=0.4\textwidth]{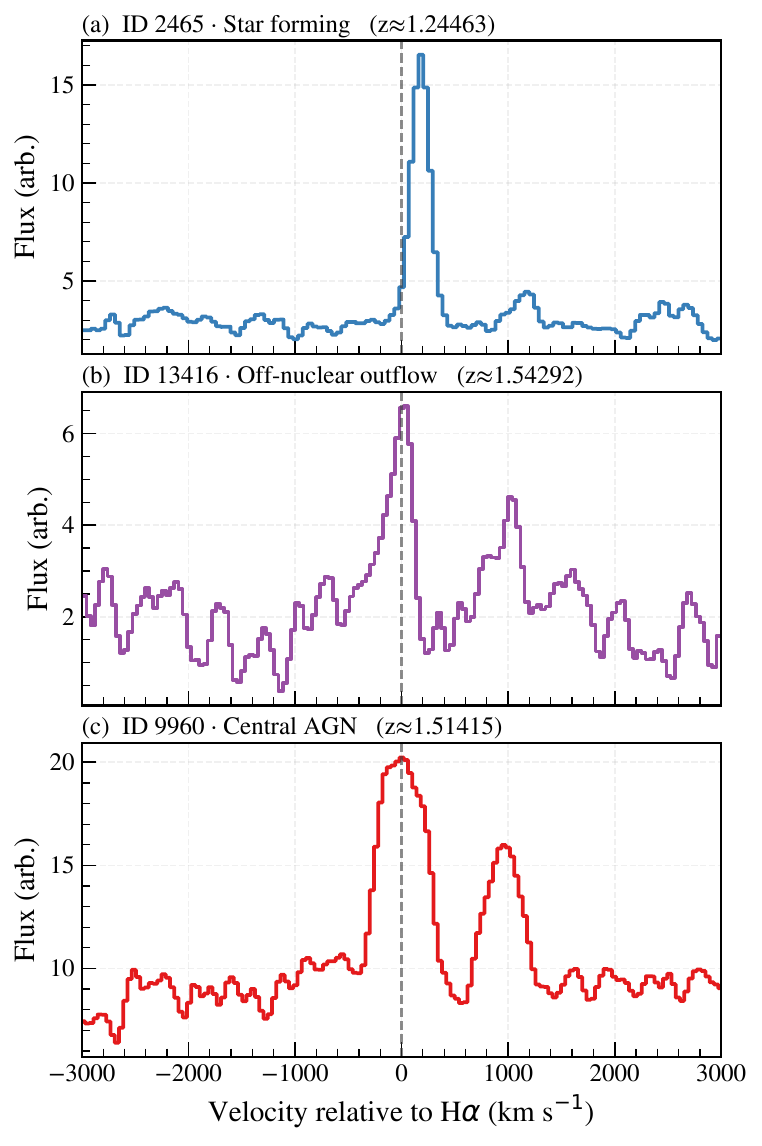}
    \caption{Rest–frame optical spectra illustrating the diversity of nebular kinematics. 
Top: emission line profile of H$\alpha$+[NII] in galaxy 2465, extracted from the central 3$\times$3 spaxels. The line profile shows a narrow, symmetric shape consistent with typical star-forming H\,\textsc{ii} regions dominated by rotating disk kinematics. 
Middle: Emission line profile for galaxy with ID 13416, extracted from a 3$\times$ 3 spaxel box, co-incident with the off-nuclear outflow location, possibly driven by supernovae. The spectral profile show a pronounced blue wing, requiring an additional kinematic component, commonly associated with bulk outflow/inflow. 
Bottom: a broad emission line feature (line width $\gtrsim$ 500 km\,s$^{-1}$) indicative of AGN-driven outflow gas in galaxy 9960. All spectra are continuum-subtracted, arbitrarily normalized, and horizontally offset to velocity = 0 for clarity.}
    \label{fig:line_profile}
\end{figure}

The kinematic maps reveal strongly disturbed gas dynamics. Regions of high
velocity dispersion ($\sigma \sim 120 - 140 \ \rm km \ s^{-1}$) extend across the system and coincide spatially with spaxels exhibiting non–star-forming ionization. These spaxels outline a biconical structure
oriented roughly perpendicular to the stellar disk. A major part of the galaxy also has  $\rm V_{RMS} \geq 300 \ km \ s^{-1}$. These values are well above those expected for purely rotation-supported disks and are inconsistent with gravitational motions alone. Additionally, we detect a secondary broad component in the emission lines, with line widths exceeding $> 500 \ \rm km \ s^{-1}$, as shown in Figure.~\ref{fig:line_profile}.
These signatures together point to the presence of high-velocity, turbulent outflows driven from the central weak AGN.

In fact, it is not unusual for local low-luminosity Seyfert galaxies to have extended ionized gas with LINER-like ratios, indicative of shocks in an outflow \citep{xia18}. 
These observed properties are also similar to the so-called ``red geysers'' identified in the MaNGA survey, which are passive, elliptical galaxies displaying large scale biconical outflows. They show similarly elevated $\sigma$ and $\rm V_{RMS}$ values (typically $\rm V_{RMS} \gtrsim 200 \ km \ s^{-1}$) and widespread LINER emission attributed to AGN-driven winds \citep{roy18, roy21a, roy21c}. Galaxy 9960 may represents a high-redshift analog of such systems exhibiting AGN-driven outflows, albeit with non negligible star formation. 

\begin{figure*}
    \centering
    \includegraphics[width=\textwidth]{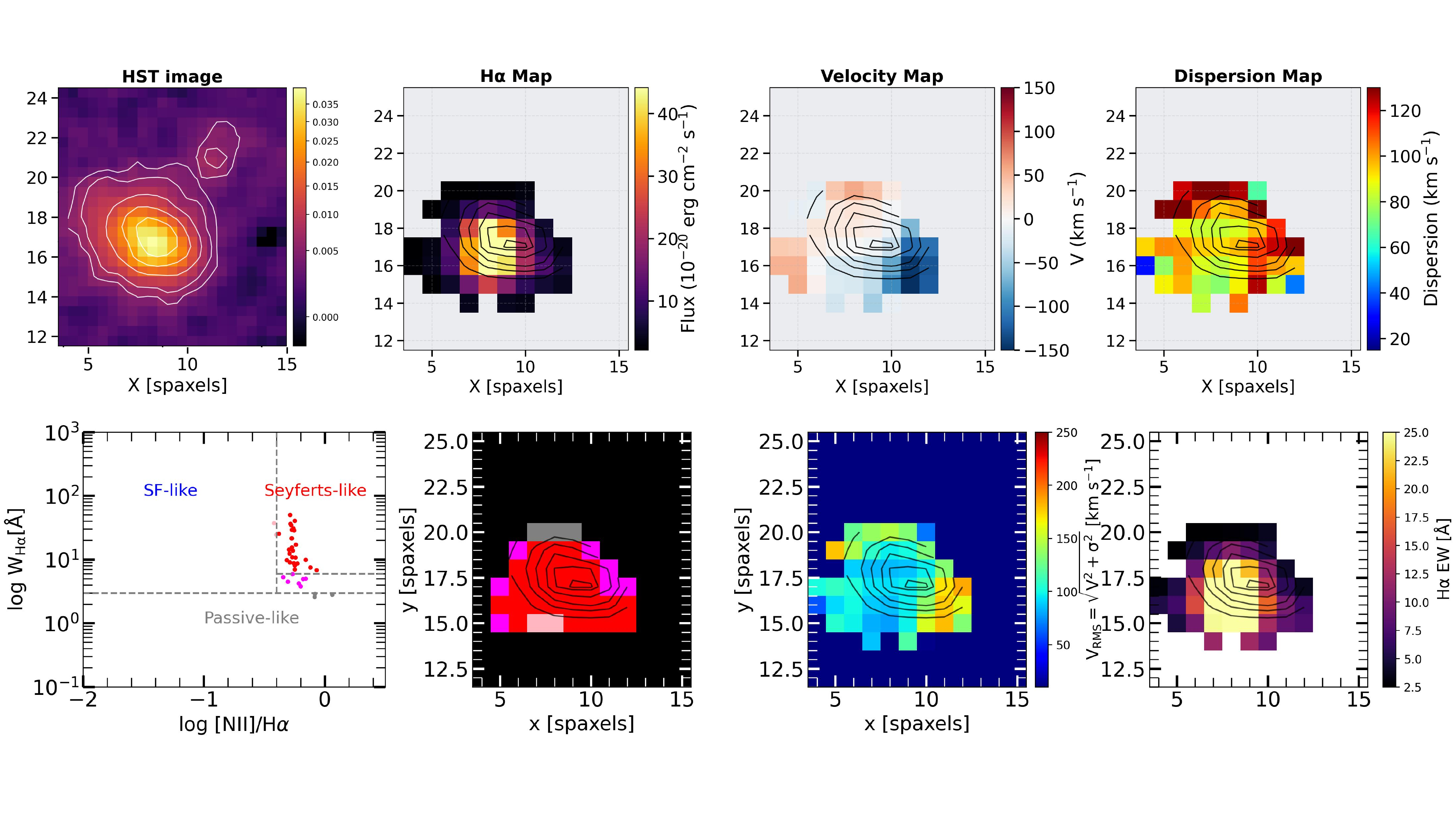}
    \caption{Spatially resolved maps for galaxy 6848. The panel layout and diagnostics shown are identical to those in Figure.~\ref{fig:2465_map}. The ionized gas is compact and dominated by non–stellar excitation. $>95\%$ of spaxels fall in the Seyfert regime of the WHAN diagram, with centrally concentrated high–excitation emission. The gas kinematics are moderately turbulent (typical $\sigma \gtrsim100~\mathrm{km \ s^{-1}}$)in the core, consistent with a compact AGN–photoionized region, with possible signature of an nuclear outflow. }
    \label{fig:6848_map}
\end{figure*}

Galaxy 6848, on the other hand, presents a compelling case with a compact ionized gas morphology dominated by non-stellar ionization mechanism.  Over 95\% of the resolved spaxels fall within the Seyfert regime in the WHAN diagram, making it the most AGN-dominated source in our sample in terms of areal coverage. Unlike more extended systems such as galaxy 9960, where shock-excited spaxels trace an extended conical pattern, here they are centrally concentrated, but spans the full extent of the detectable ionized gas.
The galaxy exhibits strong H$\alpha$ luminosity across its compact extent, suggesting a dense and highly ionized environment. Although the IFU data reveal a relatively small spatial footprint with with sufficient S/N in H$\alpha$ and [NII], HST imaging shows faint, low-surface-brightness features extending beyond the H$\alpha$ morphology. This implies that the true extent of the ionized gas is likely larger, but falls below the signal-to-noise threshold of the MSA-3D observations. Kinematically, the gas shows moderate turbulence, with velocity dispersions reaching $\sigma \gtrsim 120 \ \rm km \ s^{-1}$, and second velocity moments ($\rm V_{RMS}$) exceeding $150 \ \rm km \ s^{-1}$ in the central region.

\begin{figure*}
    \centering
    \includegraphics[width=\textwidth]{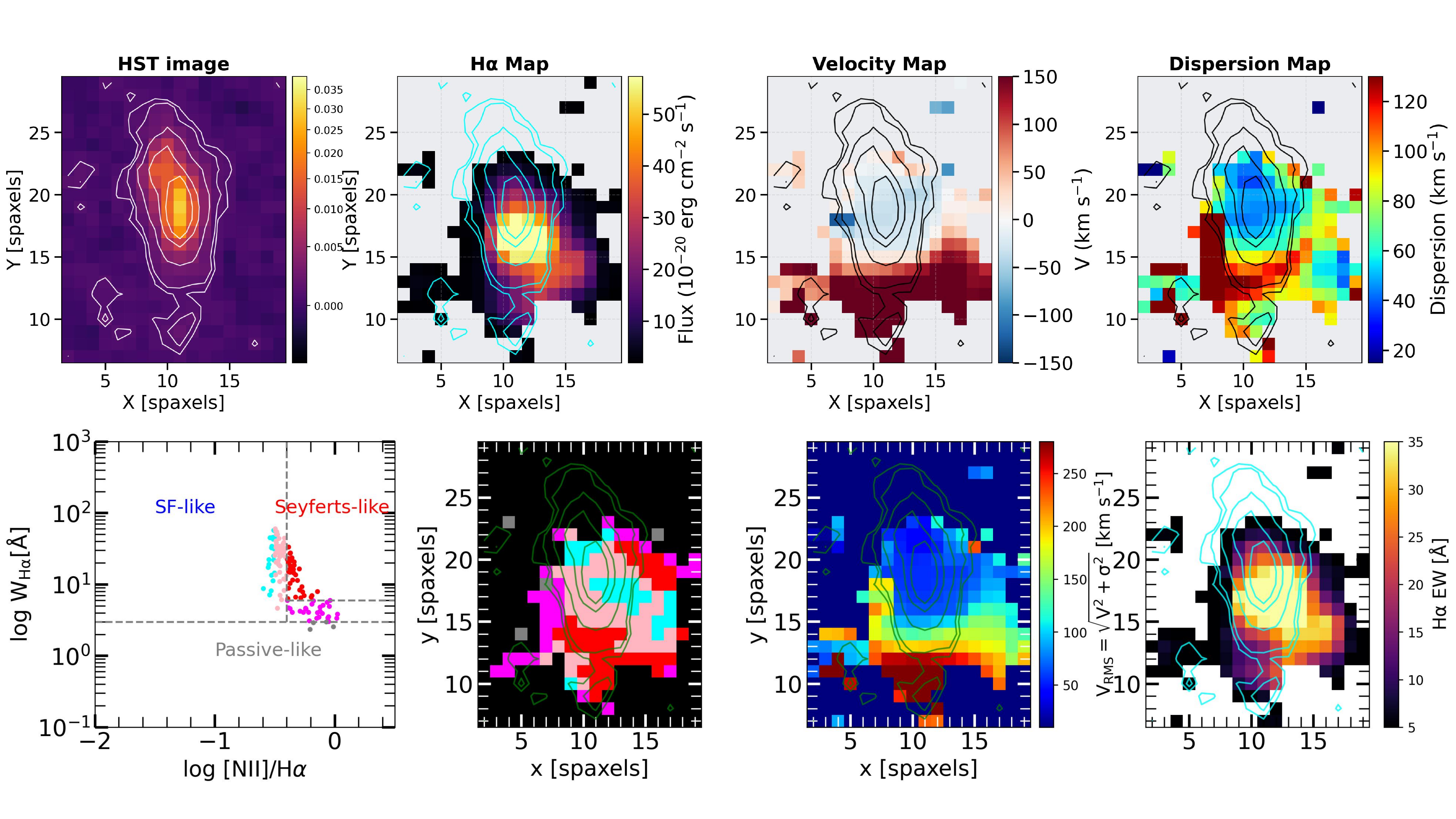}
    \caption{Spatially resolved maps for galaxy 13416.  Resolved WHAN/BPT diagnostics reveal an inverted ionization pattern with off–nuclear Seyfert/LI(N)ER spaxels encircling a more SF–like center. The gas velocity field exhibit highly asymmetric values, reaching $v>300~\mathrm{km \ s^{-1}}$ in the redshifted side, along with elevated $\sigma$. The kinematic values are consistent with a fast, off–nuclear outflow driving shock–excited emission.
 }
    \label{fig:13416_map}
\end{figure*}

\subsubsection{ID: 13416 \& 12015 -- galaxies with off-nuclear outflow or shock signatures}

Galaxy~13416 is one of two systems in our sample that falls in the
non–star-forming regime of the integrated WHAN/BPT diagnostics, with a global $\rm log([NII]/H\alpha) > -0.4$ (Figure.~\ref{fig:bpt_int}).  The spatially resolved WHAN diagram in Figure.~\ref{fig:13416_map} shows that approximately 80\% of spaxelsfall in the AGN/LINER or transitional loci (color-coded as pink, magenta or red), implying that a substantial fraction of the ionized gas is
excited by sources other than young, massive stars. Notably, the spaxels
classified as “Seyfert–like” (red) are offset from the photometric center.
Likewise, the magenta spaxels associated with shocks or LI(N)ER-like emission
are off–nuclear and trace an outer annular structure. In contrast, spaxels
classified as purely star-forming or transitional (cyan/pink) dominate the
central region. This inverted ionization pattern—harder excitation at larger
radii—suggests an extended, off–nuclear outflow, plausibly driven by star
formation or an AGN. 

The ionized–gas kinematics reinforce this picture. The velocity field is highly
asymmetric, with the redshifted side reaching  $v = +316 \rm \ km \rm \ s^{-1}$, which is
the largest amplitude in our sample and nearly three times that of the
blueshifted side.  These regions exhibit the highest velocity values across the whole sample $-$ almost 3 times larger in amplitude than even the blueshifted part of the galaxy. The velocity–dispersion map is similarly irregular and clumpy. The eastern side exhibits particularly elevated $\sigma$, where the
shock–like/AGN–like ionization is observed. Off–nuclear line profiles display
asymmetric wings (Figure~\ref{fig:line_profile}), further supporting an
outflow–driven shock interpretation.

The combination of high bulk velocities and broad dispersions is most naturally
explained by fast, outflowing gas rather than by the inflow of newly accreted
material. Such outflows could be powered by stellar feedback (e.g., supernovae
or clustered stellar winds), consistent with the off-nuclear shock-excited
spaxels. An alternative (and not mutually exclusive) scenario is a
low–luminosity or obscured AGN: while no luminous central AGN is resolved in
our MSA-3D data, a compact nucleus could evade detection at our spatial
resolution yet still drive galaxy–scale outflows that perturb the surrounding
ISM \citep{roy25}.

\begin{figure*}
    \centering
    \includegraphics[width=\textwidth]{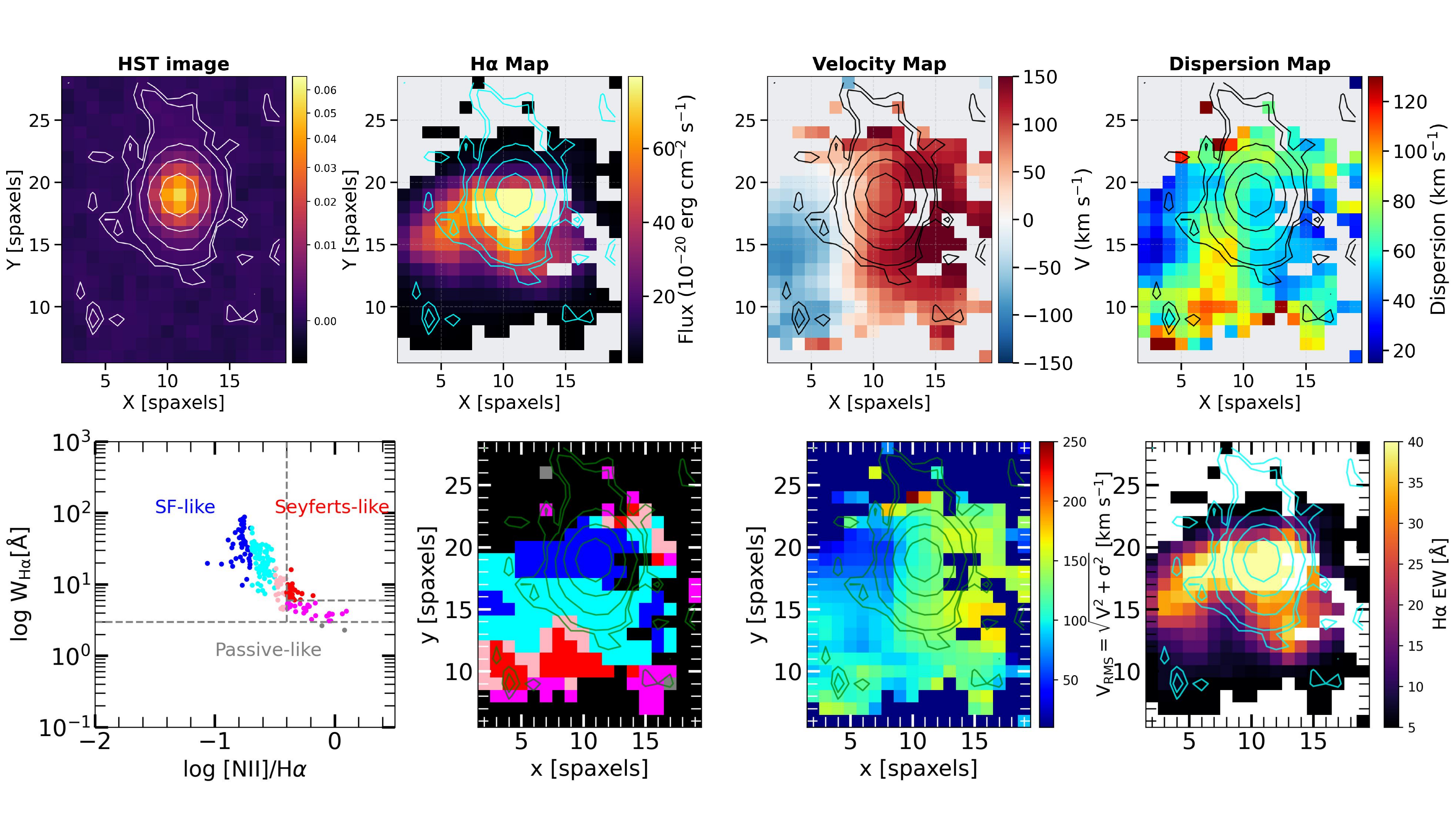}
    \caption{Resolved maps for galaxy 12015.  The gas shows a rotating disk-like velocity gradient but is kinematically misaligned with the stellar major axis (gas is aligned E–W vs. stellar photometric axis which is N–S), with 
    off–nuclear spaxels exhibiting Seyfert/LI(N)ER-like excitation along the outer edges, similar to Figure.~\ref{fig:13416_map}.}
    \label{fig:12015_map}
\end{figure*}

Figure~\ref{fig:12015_map} shows that Galaxy 12015 hosts a gas velocity gradient broadly consistent with disk-like rotation oriented east–west, yet the gaseous kinematic axis is markedly misaligned with the stellar (photometric and stellar-kinematic) major axis, which is oriented north–south. Such a gas–star kinematic misalignment is a classic signature of non-equilibrium processes and points to either inflow or outflow superposed on the underlying rotation. Nebular line ratios further reveal mixed ionization conditions: while most spaxels lie in the star-forming locus, there are notable off–nuclear regions with Seyfert/LI(N)ER-like excitation—similar to Galaxy 13416, whose incidence increases toward larger radii (red/magenta/pink loci in the WHAN diagram), indicating a contribution from non–stellar ionizing sources.

Kinematically, the system is only moderately turbulent compared to the most disturbed AGN hosts in our sample: the mean dispersion is $\sigma_{\rm mean} = 76 \pm 10 \ \rm km \ s^{-1}$, which is moderately higher than the dispersions found in typical star-forming disks at $z\sim 1$. The velocity field is relatively symmetric, and the second velocity moment remains modest with $\rm V_{RMS, avg} \approx 115 \ \rm km \ s^{-1}$, arguing against strong, galaxy-wide non-gravitational motions. The off–nuclear Seyfert/LI(N)ER-like spaxels along the outer edges are indicative of features commonly seen in nearby IFU surveys (e.g., CALIFA, MaNGA), where low-ionization emission can arise from extended shocks \citep{belfiore16,  roy21a}, photoionization by post-AGB populations \citep{yan12}, or diffuse/low-luminosity AGN illumination \citep{davies24}. In this context, Galaxy 12015 appears to be predominantly star-formation dominated, but with localized, non-stellar ionization likely powered by boundary-layer shocks from supernovae driven feedback and/or very weak AGN activity at the interfaces of turbulent structures.

\subsubsection{ID: 11843 \& 8576 -- galaxies with tentative signatures of outflow}

Galaxy 11843 and 8576 are similar because both galaxies show tentative signatures of ionized gas flows superimposed on otherwise star–formation–dominated conditions.

\begin{figure*}
    \centering
    \includegraphics[width=\textwidth]{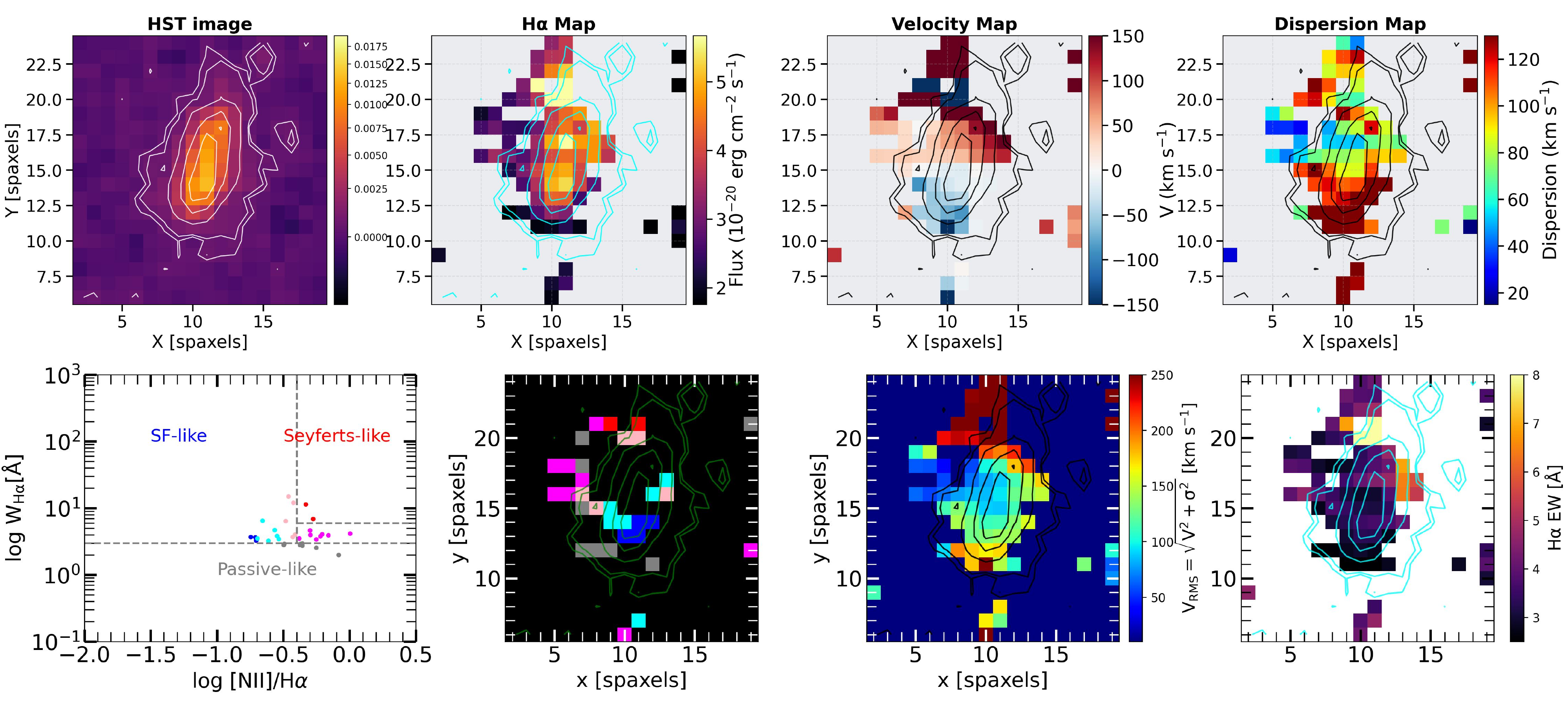}
    \caption{Maps for galaxy 11843. The panel layout and diagnostics shown are identical to those in Figure.~\ref{fig:2465_map}.}
    \label{fig:11843_map}
\end{figure*}

Galaxy 11843 is classified as composite/LINER in the integrated BPT/WHAN diagram (Figure~\ref{fig:bpt_int}), but the non–stellar excitation appears limited in strength and extent. Spatially resolved diagnostics, shown in Figure~\ref{fig:11843_map}, reveal that only a handful of spaxels have $\mathrm{S/N}>3$) in [NII] and display Seyfert- or LI(N)ER-like ratios, predominantly toward the outskirts. This suggests a lack of ionized gas and the gas may be primarily excited by evolved, LINER-like stellar populations (e.g., post-AGB). These non–SF spaxels are not randomly distributed. They coincide with localized pockets of enhanced turbulence traced by high $\sigma$ from $H\alpha$ line emission, with velocity dispersions $\sigma > 140 \ \rm km \ s^{-1}$ and second velocity moments $\rm V_{RMS} \gtrsim 300 \ \rm km \ s^{-1}$. Elevated $\sigma$ extends over a substantial north–south direction of the galaxy. Such extreme kinematics indicate fast outflows or shock–heated gas. However, their scarcity across the footprint implies that any outflow present is compact and/or faint and insufficient to produce widespread shock–excited emission—and that much of the line emission may instead reflect low–equivalent width, LINER-like excitation. Consistently, the H$\alpha$ equivalent–width map shows relatively low values across the galaxy.

In summary, the high–velocity, broad–line patches hint at outflow activity in Galaxy 11843, but the limited spatial prevalence of non–stellar excitation suggests feedback that is weak, embedded, or operating below our current sensitivity. Deeper IFU observations will be required to determine the nature and scale of feedback in this intermediate system.

Galaxy 8576 likewise exhibits tentative evidence for gas flows atop a predominantly star–forming disk. Although $\gtrsim 95\%$ of spaxels are consistent with photoionization by young stars, the gaseous velocity field is kinematically misaligned with both the photometric major axis (from HST image) and the stellar–continuum kinematic axis. This deviation from the symmetry expected for a purely rotation–supported system is commonly interpreted as a signature of radial gas motions—either inflow toward the center \citep{duckworth19, zhou21}
 or outflow escaping the disk \citep{wylezalek20, lopez19, roy21a}. In this system, the absolute kinematics are modest: $\sigma \lesssim 80 \ \rm km \ s^{-1}$, and the second velocity moment peaks at $\rm V_{RMS} \lesssim 125 \ \rm km \ s^{-1}$. The H$\alpha$ equivalent width map shows elevated EW values ($\gtrsim 30 \ $\AA) in multiple clumps near the central region, consistent with concentrated star formation, possibly driving supernovae driven winds rather than AGN-driven feedback.

Together, Galaxies 11843 and 8576 occupy a transitional regime between orderly, rotation–dominated disks and systems with clear signature of outflows. They may represent early stages of feedback or environmental influence where the dynamical impact on the ISM precedes any substantial ionization shift. As emphasized by recent work \citep{wylezalek22, davies19}, disentangling such subtle transitions requires spatially resolved kinematics and detailed ionization diagnostics to uncover the origin of disturbed gas in otherwise normal–looking galaxies.

\begin{figure*}
    \centering
    \includegraphics[width=\textwidth]{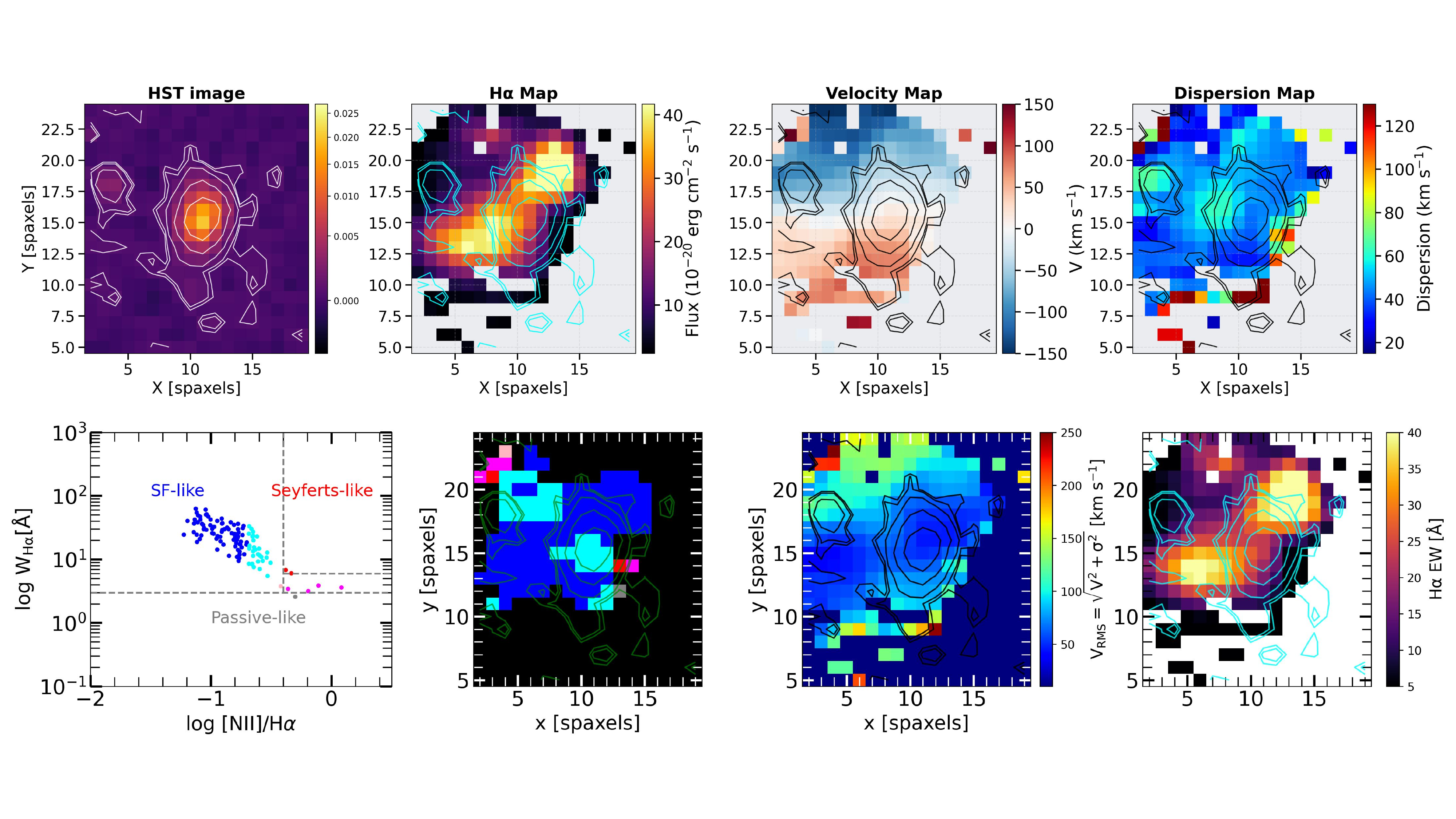}
    \caption{Maps for galaxy 8576. The panel layout and diagnostics shown are identical to those in Figure.~\ref{fig:2465_map}.}
    \label{fig:8576_map}
\end{figure*}



\section{Discussion} \label{sec:discussion}

The six AGN/outflow candidates identified here illustrate the power of spatially resolved spectroscopy to uncover weak or obscured nuclear activity that is invisible in integrated light. Although their global line ratios would classify them as ordinary, rotating star-forming disks, the resolved maps reveal compact zones of enhanced excitation, elevated velocity dispersion, and—in some cases—broad emission-line components (Figure~\ref{fig:line_profile}) indicative of non-stellar ionization. A key feature of these candidates is the presence of compact regions that exhibit subsets of spaxels crossing the \cite{kewley01} theoretical boundary for pure star formation on the BPT diagram or fall to the AGN/LINER side of the WHAN diagnostic \citep{cidfernandes11}. This points to contributions from low-luminosity AGN (LLAGN) driven photoionization, outflows and/or shock-heated gas. Similar phenomena have been reported in nearby $z<0.1$ galaxies, where spatially resolved IFU surveys \citep{roy21a, davies20, venturi21} have revealed AGN-like excitation in galaxies which are otherwise classified as purely star-forming or quenched.  In what follows, we estimate key ISM properties (electron density, ionization parameter) and approximate AGN luminosities for these candidates, and derive characteristic mass-loss rates and kinetic powers associated with their warm ionized outflows.

\subsection{Ionization and density: Inferences from Line ratio gradients of galaxies with outflows, shocks and AGNs}

Electron density ($n_e$) in these sources is derived from the [SII] $\lambda\lambda6717, 6731$ doublet ratio, which is sensitive to the density of the ionized gas. This diagnostic is only available for a subset of MSA-3D galaxies where the redshifted [SII] doublet falls within the NIRSpec G140H/F100LP wavelength range ($0.97-1.82 \mu$m). We fit the line profiles with a two-component Gaussian model and compute the flux ratio, converting it to electron density under the assumption of $T_e = 10^4\ \rm K$ \citep{osterbrock06}. The resulting $n_e$ values in the galaxies hosting AGN/outflow signatures range from 20 to 200 $\rm cm^{-3}$, consistent with typical conditions in ionized outflows.
 For comparison, the median [SII] ratio across the full MSA-3D sample is $\sim1.39$, which is consistent with the mean ratio of 1.38 observed in star-forming galaxies at $z \sim 0$ from MaNGA \citep{belfiore16}, corresponding to $n_e \sim 30\ \rm cm^{-3}$. Our AGN candidates thus span a broader and generally higher range in electron density compared to typical star-forming galaxies.

While powerful AGNs are capable of photoionizing gas over large distances, their lower-luminosity counterparts may not provide sufficient ionizing photons to explain the observed emission-line luminosities. For LLAGNs, it is crucial to determine whether the AGN alone can account for the observed nebular excitation. Hence, we estimate the ionization parameter $U$, defined as the dimensionless quantity derived from the following:

\begin{equation}
    U = \frac{Q}{4\pi r^{2}n_ec}
\end{equation}

Here, $Q$ is the ionizing photon luminosity (s$^{-1}$), $r$ is the radial distance of the line emitting gas from the AGN, and $c$ is the speed of light. To constrain $U$, we use the [OIII]$\lambda$5007/[OII]$\lambda$3727 ratio, which traces the hardness of the radiation field. Only four of the 38 galaxies in the MSA-3D sample - and none of the AGN candidates - have spectra that simultaneously cover both the [OIII] and [OII] emission lines. For these, we find [OIII]/[OII] ratios of $\sim2.20{-}3.85$, corresponding to $\log U \sim -2.8$ to $-2.5$. We assume the metallicity estimates of these galaxies from \cite{ju25}. Since these galaxies are predominantly star-forming, and AGNs are expected to exhibit harder ionizing spectra, we adopt the upper end of this range ($\log U = -2.5$) as a representative estimate for AGN-dominated environments.

Using this ionization parameter along with the mean measured electron density of $n_e \sim 110\ \rm cm^{-3}$ for our AGN sample, and a radial extent of $r \sim 5\ \rm kpc$ (based on the approximate spatial distribution of the ionized gas), we estimate the required ionizing photon rate to be $Q \approx 3 \times 10^{55}\ \rm photons\ s^{-1}$. To relate this to AGN bolometric luminosity, we adopt the average spectral energy distribution of quasars from \cite{elvis94}, where the extreme-UV slope follows $f_\nu \propto \nu^{-1.8}$. This yields a conversion of $L_{\rm AGN}/Q \approx (2{-}6) \times 10^{-11}\ \rm erg$, implying that the AGN would require a bolometric luminosity of $L_{\rm AGN} \approx 1.2 \times 10^{45}\ \rm erg\ s^{-1}$ to sustain the observed ionization level over 5 kpc.

However, this required $L_{\rm AGN}$ exceeds the AGN luminosities derived from [OIII]$\lambda$5007 or H$\alpha$ emission in the center of our sources by at least an order of magnitude. Thus, the central AGNs alone cannot supply enough ionizing photons to explain the full extent of the observed nebulae. This suggests that additional ionization sources must be contributing—such as young massive stars, shocks, or evolved stellar populations. This interpretation is further supported by resolved line ratio diagnostics, which indicate that 20–40\% of the spaxels in these galaxies lie in BPT/WHAN regions consistent with star formation. Hence, young stellar populations likely provide a substantial fraction of the ionizing radiation. 

Shocks may also play an important role. Previous work has shown that AGNs exhibiting shock-dominated ionization often display large velocity widths in forbidden lines such as [OIII] and [OII] \citep{roy21a}. Although the maximum observed FWHMs in our sample ($\lesssim 300\ \rm km\ s^{-1}$) are somewhat lower than in classical shock-dominated systems, contribution of shocks can still be non- negligible. In summary, these findings suggest that the observed nebular ionization arises from a composite of photoionization by weak AGNs, ongoing star formation, and possibly radiative shocks driven by outflows—each of which plays a varying role across spatial scales.


\subsection{Outflow rates and energetics for AGN candidates}

In this section, we estimate key physical quantities characterizing the warm ionized outflows, i.e. the ionized gas mass, mass outflow rate, kinetic power, and momentum flux, focusing on the warm $\sim10^4\ \rm K$ gas phase of the ISM.
We begin by measuring the electron density $n_e$ across the nebular extent using the [SII] $\lambda\lambda6717,6731$ doublet.  We find that the electron density ranges from 20$-$200 $\rm cm^{-3}$. 

 The ionized gas mass associated with the outflow is calculated following the prescription of \cite{cresci23}:
\begin{equation}
    \rm M_{out} = 3.2 \times 10^5 \ \frac{L_{out, H\alpha}}{10^{40} \ erg \ s^{-1}}\frac{100 \ cm^{-3}}{n_{e, out}} \ M_{\odot}
\end{equation}
where $\rm L_{out, H\alpha}$ is the H$\alpha$ luminosity associated with the outflowing gas and $n_e$ is the measured electron density. Applying this to our sources, we find ionized gas masses ranging from $3 \times 10^6$ to $2 \times 10^7\ M_{\odot}$.

Once the outflow mass is derived, the mass outflow rate for the outflowing gas can be derived by:
\begin{equation} 
    \rm \dot{M}_{out} = \frac{M_{out}v_{out}}{\Delta R_{out}}
\end{equation}
Here $\Delta R_{out}$ represents the  deprojected radial distance of the outflowing gas parcels from the host galaxy center, defined by the peak of the stellar continuum map. For our sample, the total projected extent of the outflowing gas spans $\sim5{-}8\ \rm kpc$, so we adopt $\Delta R_{out} \sim 3\ \rm kpc$, corresponding to the median radial extent. The outflow velocity is estimated as $v_{out} = \sqrt{\sigma^2 + \Delta v^2}$, where $\sigma$ is the velocity dispersion of the outflowing component and $\Delta v$ is the velocity shift relative to systemic. Using this method, we find $\dot{M}_{out} \sim 1{-}4\ M_{\odot}\ \rm yr^{-1}$ across the sample.

 The corresponding kinetic power of the outflow is given by:

\begin{equation}
    \rm \dot{KE}_{outflow}  = \frac{1}{2} \times \dot{M}_{out} v_{out}^2
\end{equation}

The resultant kinetic power (or the kinetic energy injection rates) lie between $0.7 - 1.3 \times \rm 10^{41} \ erg \ s^{-1}$. Various feedback models and simulations suggest that for AGNs to drive energetically powerful winds to inject adequate energy into the ISM, the kinetic power of the outflow must be $0.5-5\%$ of the AGN bolometric luminosity ($\rm L_{AGN}$) \citep{dimatteo05, hopkins10}. To assess this, we estimate $L_{AGN}$ from central $[{\rm OIII}]$ or H$\alpha$ luminosities measured within a $0.1''$ aperture, using empirical scaling relations \citep{heckman14}.  Our $L_{AGN}$ range from $\rm 10^{43} - 10^{44} \ \rm erg \ s^{-1}$, which implies kinetic coupling efficiencies of $\sim0.1{-}1\%$. 
This places our sources within the theoretically expected regime where AGN-driven outflows can meaningfully impact the surrounding ISM, despite their modest luminosities.

\begin{figure}
    \centering
    \includegraphics[width=0.5\textwidth]{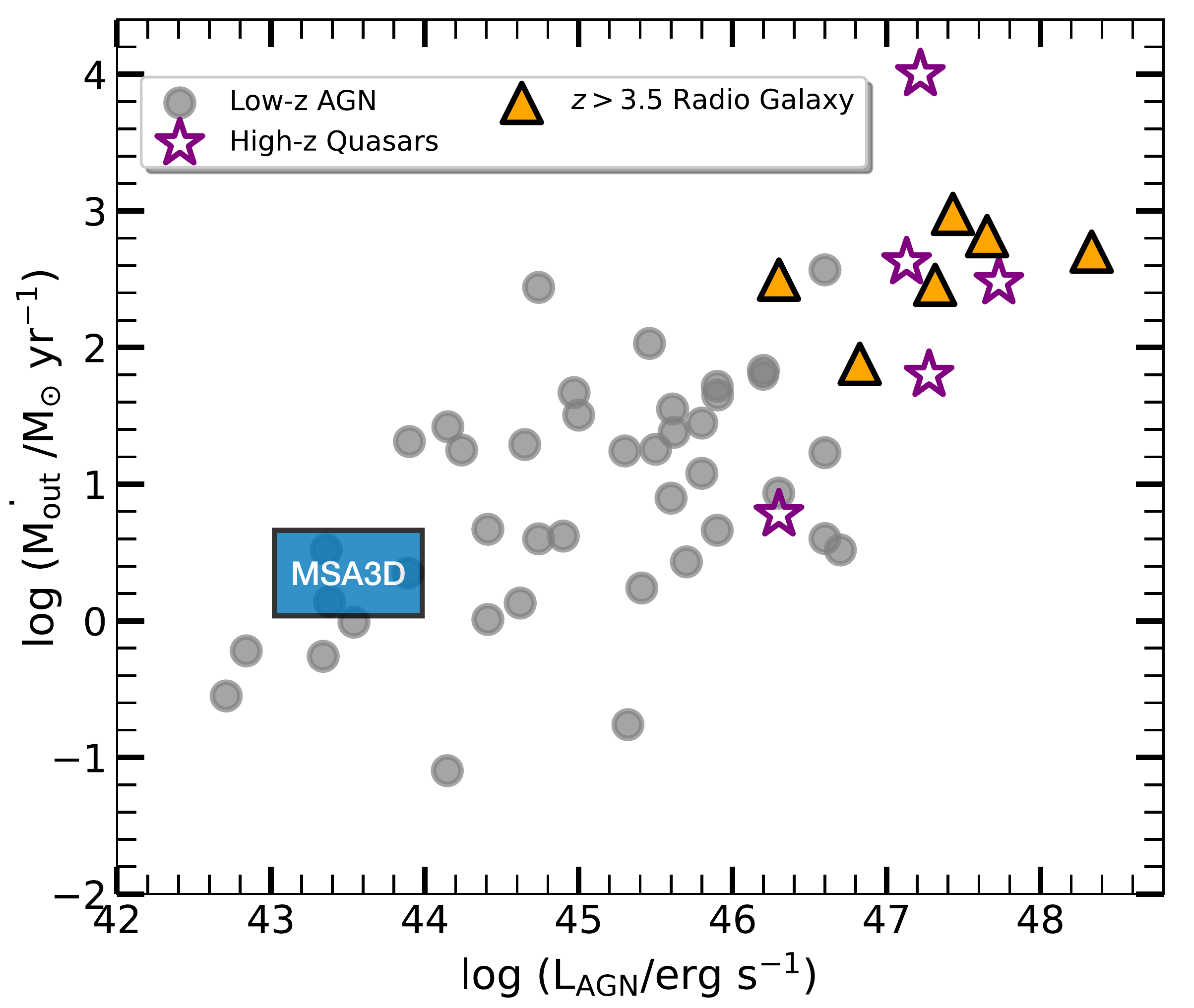}
    \caption{Integrated mass outflow rate vs. AGN bolometric luminosity for AGN sources in the MSA-3D sample (blue box) compared to sources reported in the literature using a variety of IFU/long-slit spectroscopy. Gray symbols indicate low-redshift ($z<0.1$) quasars \citep{nesvadba06, brusa16, revalski18, oliveira21, revalski21, kakkad22}, the purple stars indicate JWST/IFU observations of radio-quiet quasars at high redshifts  \citep[$1< z< 6$; ][]{cresci23, marshall23, veilleux23}, and yellow traingles represent high-z ($z>3.5$) radio galaxies \citep{roy24, wang24, wang25, roy25}. We find that the MSA-3D sources align with the broader trend between $\dot{M}_{out}$ and $L_{AGN}$, overlapping the low-luminosity distribution of AGN at $z\sim 0.1$. Their mass outflow rates constitute the lowest tail of the distribution of $z\sim 0$ AGNs.  }
    \label{fig:Mdot}
\end{figure}

To contextualize our results, Figure 15 compares the derived $\dot{M}_{out}$ and $L_{AGN}$ for our MSA-3D AGN/outflow sources to literature AGN-driven outflow measurements spanning a broad redshift range ($z \sim 0.05{-}6$). The comparison sample includes nearby low-z quasars \citep[circles:][]{nesvadba06, brusa16, revalski18, shimizu19, kakkad22}, high-z ($z > 3.5$) radio galaxies \citep[triangles: ][]{roy24, roy25, saxena24, wang24, wang25}, and luminous unobscured quasars with spatially resolved outflows \citep[stars: ][]{vayner23, cresci23, perna23,veilleux23}. Importantly, we restrict the sample to studies using directly measured $n_e$ and $v_{out}$ from spatially resolved IFU observations to ensure consistency in the derived energetics.

We find that the MSA-3D sources align with the broader trend between $\dot{M}_{out}$ and $L_{AGN}$, despite being powered by low-luminosity AGNs. Their mass outflow rates constitute the lowest tail of the distribution for the detected AGNs and quasars, suggesting that even some of the lowest lumonosity AGN activity at $z \sim 1$ can drive modest outflows, and may have similar effect as the powerful AGNs in the local Universe. This highlights the key role of low-luminosity AGNs especially at higher redshifts, and emphasize the importance of identifying this elusive population through spatially resolved spectroscopic diagnostics.

\subsection{Detection Limits and Completeness of Resolved AGN/Outflow Signatures in MSA-3D}

In the previous sections, we demonstrated that spatially resolved spectroscopy from the MSA-3D survey allows us to uncover weak AGN activity and ionized gas outflow signatures which are invisible in integrated spectra. Using emission-line diagnostics and kinematic maps, we identified six galaxies (out of 38) that show spatially extended regions indicative of LLAGN/shock photoionization and disturbed kinematics indicative of outflows. However, it is important to emphasize that our detection methodology imposes selection biases and completeness limits, which constrain the minimum AGN luminosity, outflow size, and velocity that can be reliably identified with MSA-3D.

We classify a galaxy as an AGN/outflow candidate only if it hosts at least 3 contiguous spaxels in a box lying above the AGN/LINER threshold in the WHAN diagram \citep{cidfernandes11} and exhibits elevated kinematics ($\rm V_{rms} \gtrsim 150\ km\ s^{-1}$). This ensures that isolated spaxels with uncertain classification or low signal-to-noise are excluded. However, such a criterion naturally introduces incompleteness: small-scale AGN narrow-line regions or compact outflows with projected sizes $< 5\ \rm kpc^2$ (i.e., 3 contiguous spaxels in a box at $z \sim 1$) and lower kinematic amplitudes may escape detection. So our our sample is likely missing AGNs with weaker or spatially confined signatures that fall below this threshold.

The sensitivity of the MSA-3D observations further sets a practical limit on the faintest AGN that can be detected with MSA-3D. From the measured emission line and continuum levels, the $5\sigma$ line–detection threshold corresponds to an
[OIII] $\lambda$5007 flux of $\sim 2 \times 10^{-19}\ \rm erg\ cm^{-2}\ s^{-1}$.
For the distances of our targets and standard [OIII]–to–bolometric conversions, this flux
limit implies a characteristic AGN bolometric luminosity of
$L_{\rm AGN}\approx10^{43}\ \mathrm{erg\ s^{-1}}$.
Sources with intrinsically fainter AGN or additional continuum dilution would therefore fall
below our detection sensitivity.
This imposes a conservative luminosity floor of $\rm L_{AGN} \gtrsim 10^{43}\ \rm erg\ s^{-1}$ for AGNs detected through spatially resolved line emission in the MSA-3D data.  Similarly, slow or compact outflows that do not extend over multiple spaxels or reach high velocity dispersion may also go undetected, even if physically present.
Despite these limitations, this represents a significant improvement over traditional integrated-light spectroscopy. We identify 6 candidates, but only 2 would have been identified as non star forming from traditional BPT/WHAN diagnostics from integrated spectra. Expanding MSA-3D-type surveys to deeper sensitivities, finer spatial resolution, and broader wavelength coverage will be key to uncovering the full population of faint AGNs and low-level feedback at cosmic noon.

\section{Conclusions} \label{sec:conclusion}

\noindent We present spatially resolved rest–optical spectroscopy for 38 star-forming galaxies at ($0.5<z<1.5$) from the JWST/NIRSpec MSA-3D survey, which uses slit-stepping to build IFU-like datacubes at $0.1''$ resolution. With maps of \([\mathrm{N\,II}]/\mathrm{H}\alpha\), \([\mathrm{S\,II}]/\mathrm{H}\alpha\), \([\mathrm{O\,III}]/\mathrm{H}\beta\), and \([\mathrm{N\,II}]/[\mathrm{S\,II}]\), we compare internal excitation and kinematics and place them in context with $z\sim0$ IFU results. At fixed stellar mass relative to local galaxies, our $z\sim1$ galaxies show systematically lower \([\mathrm{N\,II}]/\mathrm{H}\alpha\) and \([\mathrm{S\,II}]/\mathrm{H}\alpha\) and higher \([\mathrm{O\,III}]/\mathrm{H}\beta\), consistent with lower metallicities and harder radiation fields. Radially, \([\mathrm{N\,II}]/\mathrm{H}\alpha\) is typically flat or mildly negative, \([\mathrm{S\,II}]/\mathrm{H}\alpha\) rises with radius, \([\mathrm{O\,III}]/\mathrm{H}\beta\) is flat or slightly positive, and \([\mathrm{N\,II}]/[\mathrm{S\,II}]\) declines with radius. These trends imply only weak radial variation in O/H but a systematic change in ionization structure: the rise in [SII]/H$\alpha$ points to an increasing contribution from diffuse ionized gas and/or a lower ionization parameter at larger radii, while the  positive [OIII]/H$\beta$ is consistent with mildly declining metallicity and harder radiation fields in the outskirts. The declining [NII]/[SII] further indicates an outward decrease in N/O, as expected for inside–out chemical enrichment with delayed secondary nitrogen production.

\noindent A key result is a positive correlation between \([\mathrm{N\,II}]/\mathrm{H}\alpha\) and the velocity dispersion \(\sigma\) in kpc-scale, linking local excitation sources to turbulent kinematics. While integrated BPT/WHAN diagnostics largely place the galaxies in the star-forming regime, resolved maps reveal compact zones with AGN/Seyfert- or LINER-like excitation and elevated \(\sigma\). Six galaxies, i.e. $\sim 16\%$ of the sample, host coherent, non–star-forming ionized regions with enhanced \([\mathrm{N\,II}]/\mathrm{H}\alpha\), high \(V_{\rm RMS}=\sqrt{v^2+\sigma^2}\), and in some cases broad components, indicating weak AGN activity and/or outflows.

\noindent For these candidates, inferred warm-ionized outflow rates are modest (\(\sim1{-}4~M_\odot~\mathrm{yr}^{-1}\)) with kinetic powers \(\sim0.1{-}1\%\) of \(L_{\rm AGN}\), placing them at the low-energy tail of known AGN-driven winds yet on the same global \(\dot M_{\rm out}\)–\(L_{\rm AGN}\) trend. The survey sensitivity implies completeness for detecting AGN activity with \(L_{\rm AGN}\gtrsim10^{43}~\mathrm{erg\,s^{-1}}\) and outflows \(\gtrsim~ 2\mathrm{kpc}^2\) in size, highlighting both the potential and current limitations of detecting faint feedback at \(z\sim1\) using spatially resolved spectroscopy.

\begin{acknowledgements}

\end{acknowledgements}

\bibliography{main}
\bibliographystyle{aasjournal}
\end{document}